\documentclass[10pt,superscriptaddress,amsmath,amssymb,aps,pra,twocolumn,floatfix]{revtex4-2}

\usepackage{graphicx}
\usepackage{bm}
\usepackage[colorlinks, linkcolor=mycol, citecolor=mycol, urlcolor=mycol, breaklinks]{hyperref}
\usepackage{braket}
\usepackage{bbold}
\usepackage{bm}
\usepackage{xcolor}

\newcommand{\mi}{\mathrm{i}}

\newcommand{\id}{\mathbb{1}}
\renewcommand{\Re}{\mathrm{Re}}
\renewcommand{\Im}{\mathrm{Im}}

\definecolor{mycol}{RGB}{10,55,130}

\begin{document}

\title{Adiabatic elimination for ensembles of emitters in cavities with dissipative couplings}

\author{D.~Hagenm\"uller}
\affiliation{IPCMS (UMR 7504) and ISIS (UMR 7006), Universit\'e de Strasbourg, CNRS, 67000 Strasbourg, France}

\author{S.~Sch\"utz}
\affiliation{IPCMS (UMR 7504) and ISIS (UMR 7006), Universit\'e de Strasbourg, CNRS, 67000 Strasbourg, France}

\author{G.~Pupillo}
\affiliation{IPCMS (UMR 7504) and ISIS (UMR 7006), Universit\'e de Strasbourg, CNRS, 67000 Strasbourg, France}


\author{J.~Schachenmayer}
\affiliation{IPCMS (UMR 7504) and ISIS (UMR 7006), Universit\'e de Strasbourg, CNRS, 67000 Strasbourg, France}


\begin{abstract}
We consider an ensemble of cavity-coupled two-level emitters interacting via full (coherent and dissipative) dipole-dipole interactions. Using an adiabatic elimination procedure we derive effective equations of motion for a subsystem consisting of the cavity and a single emitter. Those equations can be used for schemes to enhance the cavity-coupling of single emitters as shown in [S.~Sch\"utz, J.~Schachenmayer, D.~Hagenm\"uller, G.~K.~Brennen, T.~Volz, V.~Sandoghdar, T.~W.~Ebbesen, C.~Genes, and G.~Pupillo, e-print arXiv:1904.08888]. We analyze limitations of effective subsystem parameters, and study how joint dissipative decay processes in the subsystem affect cavity-coupling properties of the single emitter and cavity transmission spectra.
\end{abstract}

\maketitle

In Ref.~\cite{Schutz_Ensemb_2019} we proposed a scheme to collectively enhance the coupling of a single quantum emitter ($A$) to a cavity mode via the presence of a nearby ensemble of emitters ($B$) that couple to both the cavity and $A$. Such a scheme may be useful in the context of recent experiments with molecular cavity quantum electrodynamics (QED)~\cite{Pockrand1982,Lidzey1998,Bellessa2004,dintinger2005,Shalabney2015,Chikkaraddy2016,Wang2017,Zhang2017,Wang2017}, where vibrational bands can be coupled to photons in the presence of an active molecular environment~\cite{Lather2018,Thomas_Explor_2019}. Furthermore, the scheme may find applications in quantum information processing with color centers~\cite{Park2006,Bradac2017,Astner2017,Angerer2018,Evans2018}, by collectively enhancing photon non-linearities, or may be relevant to other coupling enhancement schemes via auxiliary optical or mechanical resonators~\cite{Liu_Cohere_2014,Ramirez-Munoz_Indire_2018}.

Here, we present in detail the procedure for microscopic adiabatic elimination of the ensemble $B$ used in Ref.~\cite{Schutz_Ensemb_2019}. Our approach is an extension to the projection method given in Ref.~\cite{Reiter_Effect_2012} for the general situation where dissipative couplings between $A$ and $B$ are also present. In particular, we highlight that those dissipative couplings can also lead to effective coherent Hamiltonian terms between $A$ and the cavity. While the general effect of dissipatively engineered coherent interactions has been well studied for quantum information applications~\cite{Poyatos_Quantu_1996,Kraus_Prepar_2008,Muschik_Dissip_2011,Wang_Reserv_2013,Zanardi_Cohere_2014,Arenz_Emergi_2019}, we point out that it may also be important for cavity-coupled molecular systems. In this work we furthermore study the limitations of achievable effective parameters in the subsystem $\mathcal{S}$ consisting of $A$ and the cavity, and we observe that the effective dynamics of $\mathcal{S}$ features collective decay processes. We analyze how those collective processes modify the on-set of strong coupling between $A$ and the cavity, and how they can be responsible for an asymmetry in the cavity transmission spectrum~\cite{Canaguier-Durand_Non-Ma_2015,Neuman_Origin_2018}.

The paper is organized as follows: In Sec.~\ref{sec:full_qme}, we introduce our model and describe the quantum master equation for the full system consisting of $A$, $B$, and the cavity. In Sec.~\ref{sec:eff_qme} we derive in detail the effective master equation for the subsystem $\mathcal{S}$. We first  use an extension of the method given in~\cite{Reiter_Effect_2012}, and derive the effective master equation parameters [Sec.~\ref{sec:full_elim}]. We then show that the same effective parameters appear in a linear classical approach valid in a low excitation limit [Sec.~\ref{sec:class_elim}]. In Sec.~\ref{sec:discussion} we proceed to discuss the limitations of the effective parameters of the subsystem and their dependence on the geometry. Therefore, we analyze the case where $B$ can be reduced to an effective single emitter [Sec.~\ref{sec:eff_singlespin}] and provide analytical formulas for the modification of the parameters in this case. We then focus on the consequences of collective decay processes of $A$ and the cavity  [Sec.~\ref{sec:coll_diss}], which effectively appear after the adiabatic elimination of $B$. Finally, we provide a conclusion and an outlook in Sec.~\ref{sec:conclusion}.
\begin{figure*}[t]
\centering
\includegraphics[width=0.8\textwidth]{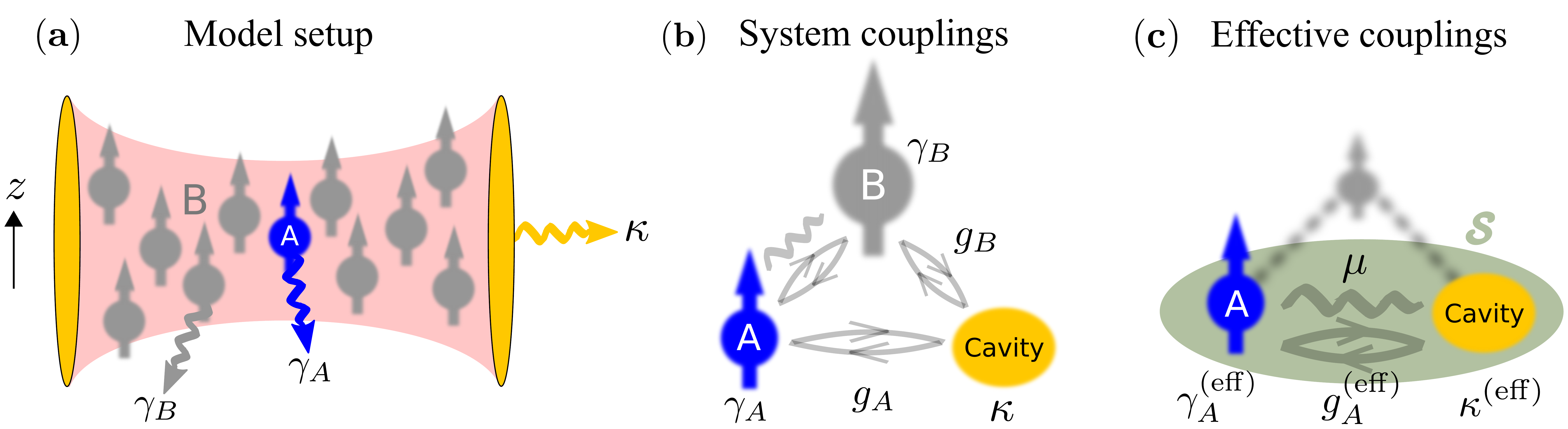}
\caption{({\bf a}) Sketch of the model setup: A single quantum emitter $A$ (decay rate $\gamma_{A}$) and an ensemble of nearby emitters $B$ (decay rates $\gamma_B$) are coupled to a cavity mode (decay rate $\kappa$). ({\bf b}) Schematics of couplings: $A$ and $B$ couple to the cavity with respective coherent Hamiltonian terms $\propto g_{A}$ and $\propto g_{B}$ (sketched as rounded arrows). The emitters interact among each others via dipole-dipole interactions with both Hamiltonian (rounded arrows) and dissipative terms (sketched as wiggly line). ({\bf c}) Schematics of couplings in the subsystem $\mathcal{S}$ after adiabatic elimination of $B$: The presence of $B$ modifies $\gamma_A \to \gamma_A^{({\rm eff})}$ and $\kappa \to \kappa^{({\rm eff})}$. Effective couplings in the subsystem comprise both coherent ($g_A \to g^{\rm (eff)}_{A}$) and dissipative couplings ($\mu$).}
\label{fig:sketch_ABcav}
\end{figure*}

\section{Model}
\label{sec:full_qme}

We consider a single two-level emitter $A$ (level spacing $\omega_{A}$, decay rate $\gamma_{A}$), an ensemble $B$ consisting of $N$ emitters (each with level spacing $\omega_{B}$, decay rate $\gamma_{B}$), and a cavity mode (frequency $\omega_{c}$, decay rate $\kappa$) [see Fig.~\ref{fig:sketch_ABcav}({\bf a}), $\hbar \equiv 1$ throughout this paper]. All emitters are coupled to the cavity and interact with each other via dipole-dipole interactions (the dipole direction is chosen along the $z$ axis). In a frame rotating with $\omega_A$, the full quantum master equation for the density matrix $\rho$ of the system can be written as
\begin{align}
\partial_t \rho = - \mi [ H_0 + H_{\text{TC}} + H_{\text{DD}} , \rho ] + \mathcal{L}\rho.
\label{full_qme}
\end{align}

Here, the Hamiltonian parts include the bare system energies,
\begin{align*}
H_0 
= \Delta_c a^{\dagger} a 
+ \sum_{j=1}^{N} \Delta_{B} \sigma_j^+ \sigma_j^-
\end{align*} 
 with $\Delta_c = \omega_{c}-\omega_{A}$ and $\Delta_B = \omega_{B}-\omega_{A}$. The bosonic operators $a$ and $a^{\dagger}$ annihilate and create a photon in the cavity, while $\sigma^{\pm}_{A}$ and $\sigma^{\pm}_{j}$ are the spin ladder operators for $A$ and for the $j^{\textrm{th}}$ emitter of the ensemble $B$, respectively. The emitter-cavity interaction is governed by a Tavis-Cummings Hamiltonian~\cite{Jaynes_Compar_1963,Tavis_Exact_1968},
\begin{align*}
    H_{\text{TC}} = a^{\dagger} \Big( g_A \sigma_A^- + \sum_{j=1}^{N} g_{j} \sigma_j^- \Big) + \textrm{h.c.}
\end{align*}
with respective coupling strengths $g_{A}$ and $g_{j}$. The dipole-dipole interaction Hamiltonian is given by~\cite{Lehmberg1970_Radia}
\begin{align*}
H_{\text{DD}} &= 
\sum_{j=1}^{N} \Omega_{j A} \big( \sigma_{j}^+ \sigma_A^- + \sigma_A^+ \sigma_{j}^- \big)
+ \sum_{j \neq \ell}^{N} \Omega_{j \ell} \sigma_{j}^+ \sigma_{\ell}^{-}, \nonumber
\end{align*}
with $\Omega_{j A}$ denoting the coupling strengths between $A$ and the $j^{\rm th}$ spin of $B$, and $\Omega_{j\ell}$ the coupling strengths between pairs of emitters within $B$. 

Dissipation in Eq.~(\ref{full_qme}) is described by the super-operator
\begin{align*}
\mathcal{L} \rho 
= - \kappa \mathcal{D} (a^{\dagger},a) \rho - \gamma_A \mathcal{D} (\sigma_A^+, \sigma_A^-) \rho 
+ \mathcal{L}_{\rm BB} \rho + \mathcal{L}_{\rm AB} \rho
\end{align*}
with $\mathcal{D} (x,y) \rho = [x, y \rho] + [\rho x, y]$. The dissipator given by $\mathcal{L}_{\rm BB} \rho = - \sum_{j,\ell= 1}^{N} \gamma_{j \ell} \mathcal{D} (\sigma_{j}^+,\sigma_{\ell}^-) \rho$ describes collective decay processes within the $B$ ensemble, and $\gamma_{jj} = \gamma_B$. Additionally,
\begin{align*}
\mathcal{L}_{\rm AB} \rho = - \sum_{j=1}^{N} \gamma_{j A} \Big( \mathcal{D} (\sigma_{j}^+ ,\sigma_A^-) \rho + \mathcal{D} (\sigma_{A}^+,\sigma_j^-) \rho \Big)
\end{align*}
describes dissipative couplings between $A$ and $B$ [see Fig.~\ref{fig:sketch_ABcav}({\bf b}) for a sketch of the couplings]. 

In this work we are interested in deriving effective equations of motions for the density matrix of the subsystem $\mathcal{S}$ consisting of $A$ and the cavity after adiabatically eliminating $B$ [see Fig.~\ref{fig:sketch_ABcav}({\bf c})]. As we show below, besides energy shifts, those equations feature modified decay rates ($\gamma_A \to \gamma_A^{({\rm eff})}$ and $\kappa \to \kappa^{({\rm eff})}$), as well as modified couplings between $A$ and the cavity. The latter consist of both coherent Jaynes-Cummings type couplings (with an effective coupling strength $g_A \to g^{\rm (eff)}_{A}$) as well as collective dissipation terms with rate $\mu$.

\section{Derivation of effective quantum master equation}
\label{sec:eff_qme}

In this section, we derive the effective quantum master equation obtained when the emitter ensemble $B$ is adiabatically eliminated, which leads to modified physical parameters for the subsystem $\mathcal{S}$ [Fig.~\ref{fig:sketch_ABcav}({\bf c})]. The main results are the effective master equation parameters provided in Sec.~\ref{sec:eff_para}. The derivation of the parameters is split into two parts, in Sec.~\ref{sec:full_elim} we utilize a full adiabatic elimination procedure similar to that of Ref.~\cite{Reiter_Effect_2012}. In Sec.~\ref{sec:class_elim} we show that the same effective parameters also appear in a classical linear theory for low spin excitations.
 
 \subsection{Full adiabatic elimination procedure}
 \label{sec:full_elim}
 
Our derivation of the effective master equation in this section relies on elimination techniques (see e.g.~\cite{Zwanzig1960,Bonifacio1971,Schuetz2013,PhysRevB.97.205303}) and is based on a projection method as given in Ref.~\cite{Reiter_Effect_2012}, which we extend to the general case where dissipative dipole-dipole couplings between $A$ and $B$ (i.e.~$\mathcal{S}$ and $B$) are also present. This method relies on the projection of the density operator onto the ground-state manifold of $B$ (Sec.~\ref{sec:proman}), suitable decompositions of the master equation and the density operator (Sec.~\ref{sec:decomp_feca} and Sec.~\ref{sec:decomp_densi}), followed by second-order perturbation theory in the interaction between $\mathcal{S}$ and $B$ (Sec.~\ref{sec:pert_2ord}) and time-scale separation (Sec.~\ref{sec:tim_scal}). The main result of the adiabatic elimination are the effective parameters for the subsystem $\mathcal{S}$ master equation and are provided in Sec.~\ref{sec:eff_para}. 

\subsubsection{Projection onto the ground-state manifold}
\label{sec:proman}

We assume that the set $\{ |s_i \rangle \}$ forms a basis for the subsystem $\mathcal{S}$, while $\{ |b_i \rangle \}$ is a basis for the interacting ensemble $B$. The density operator $\rho$ of the full system can be written as $\rho = \sum_{ijkl} \rho_{ij;kl} |s_i \rangle \langle s_j| \otimes|b_k \rangle \langle b_l |$ with matrix elements $\rho_{ij;kl} = \langle s_i,b_k | \rho |s_j, b_l \rangle $ and $|s_j, b_l \rangle = |s_j \rangle \otimes |b_l \rangle $. We are interested in the time evolution of the reduced density operator $\rho^{\text{eff}} = \sum_{ij}  \rho^{\text{eff}}_{ij} |s_i \rangle \langle s_j |$ of $\mathcal{S}$ obtained by taking the partial trace $\text{Tr}_{B} \rho = \sum_m \langle b_m | \rho | b_m \rangle$ over the $B$ ensemble, where $\rho^{\text{eff}}_{ij} = \sum_m \rho_{ij;mm}$ involves a sum over all possible basis states of $B$. 

We assume that the state $| g \rangle $ with {\it all} spins of $B$ in their ground states mainly contributes to the latter sum, and therefore introduce the super-operator $P \bullet = |g \rangle \langle g | \bullet |g \rangle \langle g |$ as a projector onto the elements of interest of the density operator $\rho$ with $P \rho = \rho_{gg} |g \rangle \langle g |$ and $\rho_{gg} = \langle g | \rho |g \rangle$. In the following, we derive an equation for the time evolution of $v = P \rho$. Under the assumption that the spins of $B$ remain close to their ground states, we derive the effective time evolution for the reduced density operator $\rho^{\text{eff}}$ of $\mathcal{S}$ with $\rho^{\text{eff}} \approx \rho_{gg}$. 

\subsubsection{Decomposition of the master equation}
\label{sec:decomp_feca}

We start by decomposing the full quantum master equation (\ref{full_qme}) for the total density operator $\rho$ as
\begin{align}
\partial_t \rho = \mathfrak{L} \rho = ( \mathcal{L}_{S} + \mathcal{L}_{B} + J + \mathcal{L}_{\rm{int}}) \rho.
\label{eq:decompL}
\end{align}

The first term $$\mathcal{L}_{S} \rho = - \mi [H_S , \rho] + \mathcal{L}_{A} \rho + \mathcal{L}_{c} \rho$$ is associated to the dynamics in $\mathcal{S}$, which includes the coherent evolution governed by the Hamiltonian $H_S = \Delta_c a^{\dagger} a + g_A \big( \sigma_A^+ a + a^{\dagger} \sigma_A^- \big)$, as well as a coupling to the environment associated to the terms $\mathcal{L}_{A} \rho = - \gamma_A (\sigma_A^+ \sigma_A^- \rho + \rho \sigma_A^+ \sigma_A^- - 2 \sigma_A^- \rho \sigma_A^+)$ for the spin $A$ and $\mathcal{L}_{c} \rho = - \kappa (a^{\dagger} a \rho + \rho a^{\dagger} a - 2 a \rho a^{\dagger} )$ for the cavity mode. 

The second term reads 
\begin{align*}
\mathcal{L}_B \rho = -\mi \vec{\sigma}^{\rm T}_{+} {\bf M} \vec{\sigma}_{-} \rho + \mi \rho \vec{\sigma}^{\rm T}_{+} {\bf M}^{*} \vec{\sigma}_{-}.
\end{align*}
Here, we used vector and matrix notations ($^\text{T}$ denotes the transpose operation) with 
\begin{align}
({\bf M})_{j \ell}=\big(\Delta_B - \mi \gamma_B \big) \delta_{j \ell} + (1-\delta_{j \ell}) (\Omega_{j \ell} - \mi \gamma_{j\ell}), 
\label{eq:matrix_M}
\end{align}
and the spin ladder operators of $B$ $(\vec{\sigma}_{\pm})_{j}=\sigma^{\pm}_{j}$. The matrix ${\bf M}$ describes the internal dynamics of the $B$ ensemble, namely the free evolution $\sim \Delta_{B}$ of each spin of $B$, their couplings to the environment $\sim \gamma_{B}$, as well as their mutual interactions due to coherent ($\sim \Omega_{j\ell}$) and incoherent ($\sim \gamma_{j\ell}$) dipole-dipole interactions. 

The third term in Eq.~(\ref{eq:decompL}) reads 
$$J \rho = 2 \vec{\sigma}^{\rm T}_{-} \bm{\gamma} \rho \vec{\sigma}_{+},$$ 
and includes both individual (diagonal) and correlated (off-diagonal) terms for the ensemble $B$ with $(\bm{\gamma})_{j \ell} = \gamma_{j \ell}$. 

The last term in Eq.~(\ref{eq:decompL}) is given by
\begin{align}
\mathcal{L}_{\rm{int}} \rho = &-\mi \left[\left(a \vec{G}^{\rm T} + \sigma^{-}_{A}\vec{V}^{\rm T} \right) \vec{\sigma}_{+} + \left(a^{\dagger} \vec{G}^{\rm T} + \sigma^{+}_{A}\vec{V}^{\rm T} \right) \vec{\sigma}_{-} \right] \rho \nonumber \\
&+\mi \rho \big[\big(a \vec{G}^{\rm T} + \sigma^{-}_{A}\vec{V}^{*{\rm T}} \big) \vec{\sigma}_{+} + \big(a^{\dagger} \vec{G}^{\rm T} + \sigma^{+}_{A}\vec{V}^{*{\rm T}} \big) \vec{\sigma}_{-} \big] \nonumber \\
&+ 2 \left(\sigma^{-}_{A} \rho (\vec{\gamma}^{\rm T} \vec{\sigma}_{+}) + (\vec{\gamma}^{\rm T} \vec{\sigma}_{-}) \rho \sigma^{+}_{A} \right),
\label{L_int_eq11}
\end{align}
and describes the coupling of $B$ to $A$ and to the cavity mode. Here, we introduced the vector notations $(\vec{G})_{j} = g_{j}$ and $\vec{V}=\vec{\Omega}-\mi \vec{\gamma}$ with $(\vec{\Omega})_{j}=\Omega_{j A}$ and $(\vec{\gamma})_{j} = \gamma_{j A}$. While coherent spin-cavity and spin-spin couplings are encoded in $\vec{G}$ and $\vec{\Omega}$, respectively, the terms $\propto \vec{\gamma}$ describe the dissipative part of the spin-spin coupling between $A$ and $B$.
The latter constitute a dissipative coupling between $\mathcal{S}$ and $B$.

\subsubsection{Decomposition of the density operator}
\label{sec:decomp_densi}

With the convenient form of Eq.~(\ref{eq:decompL}), we can now decompose the total density operator as 
$$\rho = (P + Q) \rho = v + w$$ 
with $w = Q \rho = (1 - P) \rho$, and derive an equation of motion for the projection $v=\rho_{gg}| g \rangle \langle g |$ of the density operator onto the ground-state manifold of $B$. The projectors $P$ and $Q$ fulfill the relations $P^2 \rho = P \rho$, $Q^2\rho = Q \rho$, and $PQ \rho = QP \rho = 0$. Using Eq.~(\ref{eq:decompL}), the time evolution of the operators $v$ and $w$ is given by $\partial_t v = P (\partial_t \rho) = P \mathfrak{L} P \rho + P \mathfrak{L} Q \rho$ and $\partial_t w = Q (\partial_t \rho) = Q \mathfrak{L} P \rho + Q \mathfrak{L} Q \rho$. One can show that among all possibilities stemming from the different contributions to $\mathfrak{L}$, the only non-vanishing terms are 
\begin{alignat}{2}
P \mathfrak{L} P &= P \mathcal{L}_{S} P &\qquad P \mathfrak{L} Q &= P J Q + P \mathcal{L}_{\rm int} Q \label{eq_contrib} \\
Q \mathfrak{L} P &= Q \mathcal{L}_{\rm int} P  & Q \mathfrak{L} Q &= Q (\mathcal{L}_{S} + \mathcal{L}_B + J + \mathcal{L}_{\rm{int}}) Q. \nonumber
\end{alignat}

From now on, we assume
\begin{align*}
w = Q \rho &\simeq \sum_{j} \langle g | \rho | e_{j} \rangle | g \rangle \langle e_{j} | + \sum_{j} \langle e_{j} | \rho | g \rangle | e_{j} \rangle \langle g | \\
&+ \sum_{j,l} \langle e_{j} | \rho | e_{l} \rangle | e_{j} \rangle \langle e_{l} |,
\end{align*}
i.e.~we restrict the following calculations to the single-excitation subspace where the $B$ ensemble contains at most one excitation, consistently with the assumption that the spins of $B$ remain close to their ground states. Here, $|e_{j} \rangle$ denotes the state where the $j^{\rm th}$ spin of $B$ is in its excited state, while the others are in their ground states. This approximation allows to discard the term $QJQ \sim 0$ in Eq.~(\ref{eq_contrib}). In total, the equations of motion of the projected density operators then read
\begin{align}
\partial_t v &= P \mathcal{L}_{S} v + P ( J + \mathcal{L}_{\rm int} ) w 
\label{eq_eff},\\
\partial_t w &= Q (\mathcal{L}_B + \mathcal{L}_S) w + Q \mathcal{L}_{\rm int} v + Q \mathcal{L}_{\rm int} w. 
\label{eq_eff_fin}
\end{align}

Introducing the operator $\mathcal{L}_0= \mathcal{L}_S + \mathcal{L}_B$ that describes the free evolution of the system, the formal solution of Eq.~(\ref{eq_eff_fin}) is
\begin{align}
w (t) 
&= e^{Q \mathcal{L}_0 (t-t_0)} w(t_0)  \nonumber \\
&+ \int_{t_0}^{t} d \tau e^{Q \mathcal{L}_0 (t-\tau)} \big[ Q \mathcal{L}_{\rm int} v(\tau) + Q \mathcal{L}_{\rm int} w(\tau) \big], 
\label{eq_w_order}
\end{align}
which we now insert into Eq.~(\ref{eq_eff}). This yields
\begin{align}
\partial_t v 
&= P \mathcal{L}_S v + P (J+ \mathcal{L}_{\rm int})
\int_{t_0}^{t} d \tau e^{Q \mathcal{L}_0 (t-\tau)} Q \mathcal{L}_{\rm int} v (\tau) \nonumber \\
& +P (J+ \mathcal{L}_{\rm int}) \int_{t_0}^{t} d \tau e^{Q \mathcal{L}_0 (t-\tau)} Q  \mathcal{L}_{\rm int} \nonumber \\ 
&\times \int_{t_0}^{\tau} d \tau' e^{Q \mathcal{L}_0 (\tau-\tau')} \big[ Q \mathcal{L}_{\rm int} v(\tau') + Q \mathcal{L}_{\rm int} w(\tau') \big],
\label{eq_four} 
\end{align}
assuming that the $B$ ensemble is initially in its ground state ($w(t_0)=0$). 

Equation (\ref{eq_four}) is the desired equation of motion for the projection $v$ of the density operator onto the ground-state manifold of $B$. In the following sections, we evaluate the different contributions entering this equation using the definitions of Sec.~\ref{sec:decomp_feca}, as well as a perturbative expansion in $\mathcal{L}_{\rm{int}}$ which describes the couplings of $B$ to $A$ and to the cavity mode.

\subsubsection{Perturbation to second order in \texorpdfstring{$\mathcal{L}_{\rm int}$}{}}
\label{sec:pert_2ord}

We now consider all possible processes up to second order in $\mathcal{L}_{\rm int}$. This procedure consists of a perturbative treatment of the interaction between $A$ and the cavity mode with the quasi-modes of the interacting $B$ ensemble. It is justified when the coupling strengths of these quasi-modes to the subsystem $\mathcal{S}$ is sufficiently small compared to their (far-detuned) eigenfrequencies or dissipation rates (see e.g.~Appendix~\ref{sec:validity_class_elim}), which ensures that the spins of $B$ remain weakly excited. Up to second order in $\mathcal{L}_{\rm int}$, Eq.~(\ref{eq_four}) provides
\begin{align}
\partial_t v &= P \mathcal{L}_S v + P \mathcal{L}_{\rm int} \int_{0}^{t-t_0} d \tau e^{Q \mathcal{L}_0 \tau} Q \mathcal{L}_{\rm int} v (t-\tau) \nonumber \\ 
& +P J \int_{0}^{t-t_0} d \tau e^{Q \mathcal{L}_0 \tau} Q \mathcal{L}_{\rm int} \nonumber \\
&\times
\int_{0}^{t-t_{0}-\tau} d \tau' e^{Q \mathcal{L}_0 \tau'} Q \mathcal{L}_{\rm int} v(t-\tau-\tau').
\label{fina_equ_ad}
\end{align}

 Note that we have neglected the term $Q \mathcal{L}_{\rm int} w(\tau')$ in Eq.~\eqref{eq_four} since its contribution is at least of order $\mathcal{O} (\mathcal{L}_{\rm int}^3)$ [see Eq.~(\ref{eq_w_order})]. Furthermore, our truncation is also consistent with neglecting the term $QJQ$ in Eq.~(\ref{eq_contrib}) as we did before since the latter would provide contributions of higher order in $\mathcal{L}_{\rm int}$. In order to calculate the different contributions entering Eq.~(\ref{fina_equ_ad}), it is convenient to use a spectral decomposition of the $N\times N$ non-hermitian, complex symmetric matrix ${\bf M}$ in Eq.~\eqref{eq:matrix_M}. Assuming that the latter can be diagonalized~\cite{Horn2012}, we write ${\bf B}\equiv -\mi {\bf M}$ and ${\bf B}= \sum_{j}\lambda_{j} \vec{x}_{j}\vec{x}^{\rm T}_{j}$, where $\lambda_{j}$ and $\vec{x}_{j}$ ($j=1,\cdots,N$) denote the complex eigenvalues and eigenvectors of ${\bf B}$, respectively. The eigenvectors satisfy the completeness relation $\sum_{j}\vec{x}_{j} \vec{x}^{\rm T}_{j}= {\bm 1}$ and form an orthogonal basis with respect to the inner product $\vec{x}^{\rm T}_{j} \vec{x}_{\ell}=\delta_{j,\ell}$. 

\subsubsection{Integration using time-scale separation}
\label{sec:tim_scal}

We can now proceed with the integration of the different terms entering Eq.~(\ref{fina_equ_ad}), and first calculate quantities of the type $\mathcal{L}_{\rm int} e^{Q \mathcal{L}_0 \tau}Q \mathcal{L}_{\rm int}v(t-\tau)$. Using the definitions Eq.~(\ref{L_int_eq11}) and $v (t)= \rho_{gg} (t)| g \rangle \langle g |$, as well as the completeness relation, we obtain 
\begin{align*}
\mathcal{L}_{\rm int}v(t-\tau) 
= -\mi \sum_{j} V^{\uparrow}_j \rho_{gg} (t-\tau) | x_j \rangle \langle g | + \textrm{h.c.},
\end{align*}
where $V^{\uparrow}_j = \big(a \vec{G}^{\rm T} + \sigma^{-}_{A} \vec{V}^{\rm T} \big) \vec{x}_{j}$ describes the action onto the subsystem $\mathcal{S}$ when an excitation $| x_j \rangle = \big( \vec{x}^{\rm T}_{j} \vec{\sigma}_{+} \big) | g \rangle $ for the $j$-th eigenmode of $B$ is created.

Under the assumption that the dynamics of the subsystem $\mathcal{S}$ is slow compared to the internal dynamics of $B$, one can use a Markov-type approximation $\rho_{gg} (t-\tau) \approx \rho_{gg} (t)$ in the integrand of Eq.~\eqref{fina_equ_ad}, together with a Taylor expansion to zeroth order in $\mathcal{L}_S \tau$ of the operator $\exp [Q \mathcal{L}_S \tau ]Q$, such that $\exp [Q (\mathcal{L}_B + \mathcal{L}_S) \tau ] Q\approx \exp [ Q \mathcal{L}_B \tau ]Q$. Using the relation $\mathcal{L}_B | x_j \rangle \langle g |= \lambda_{j} | x_j \rangle \langle g |$, we obtain
\begin{align*}
e^{Q \mathcal{L}_0 \tau} Q \mathcal{L}_{\rm int} v (t - \tau) 
 \approx
-\mi \sum_j e^{\lambda_j \tau} V^{\uparrow}_j \rho_{gg} (t) | x_j \rangle \langle g | + \textrm{h.c.} 
\nonumber 
\end{align*} 
Now applying $\mathcal{L}_{\rm int}$ on the previous expression (while restricting ourselves to the single-excitation subspace) leads to
\begin{align*}
&\mathcal{L}_{\rm int} Q
\bigg( -\mi \sum_j e^{\lambda_j \tau} V^{\uparrow}_j \rho_{gg} (t) | x_j \rangle \langle g | + \textrm{h.c.} \bigg) \\
= &- \left( a^{\dagger} \vec{G}^{\rm T} + \sigma_A^+ \vec{V}^{\rm T} \right) \vec{\sigma}_{-} \sum_j e^{\lambda_j \tau} V^{\uparrow}_j \rho_{gg} (t) | x_j \rangle \langle g | + \textrm{h.c.} \\
&-2 \mi \big(\vec{\gamma}^{\rm T} \vec{\sigma}_{-} \big) \sum_j e^{\lambda_j \tau} V^{\uparrow}_j \rho_{gg} (t) | x_j \rangle \langle g | \sigma^{+}_{A} + \textrm{h.c.} \\
&+ \sum_j e^{\lambda_j \tau} V^{\uparrow}_j \rho_{gg} (t) | x_j \rangle \langle g | \left( a^{\dagger} \vec{G}^{\rm T} + \sigma_A^+ (\vec{V}^{*})^{\rm T} \right) \vec{\sigma}_{-} + \textrm{h.c.}
\end{align*}

 Integration of the previous expression according to Eq.~\eqref{fina_equ_ad}
provides terms proportional to $1- e^{\lambda_{j} (t-t_{0})} \approx 1$. This approximation is valid in the limit of large $t-t_0$ when $\Re[\lambda_j] < 0$ (e.g.~$\Re [\lambda_j] = - \gamma_B$ for a single spin $B$). More generally, it is justified when the expression is averaged on a coarse-grained time scale $\Delta t$ that fulfills $|\lambda_j|^{-1} \ll \Delta t \ll \tau_s$ with $\tau_s^{-1}$ the typical rate for the dynamics of subsystem $\mathcal{S}$. 

We now use the completeness relation once again, as well as the relations $P \big( \vec{x}_{j}^{\rm T} \vec{\sigma}_{-} \big) \big( \vec{x}_{\ell}^{\rm T} \vec{\sigma}_{+} \big) | g\rangle \langle g |=\delta_{j\ell} | g\rangle \langle g |$ and $\sum_j \vec{x}_{j}\vec{x}_{j}^{\rm T} / \lambda_{j} = {\bf B}^{-1}$, in such a way that the second term in the first line of Eq.~(\ref{fina_equ_ad}) becomes
\begin{align}
& P \mathcal{L}_{\rm int} \int_{0}^{t-t_0} d \tau e^{Q \mathcal{L}_0 \tau} Q \mathcal{L}_{\rm int} v(t-\tau) \label{secon_termm} \\
\approx \mi \Big( &[\vec{G}^{\rm T} {\bf M}^{-1} \vec{G}] a^{\dagger} a + [\vec{G}^{\rm T} {\bf M}^{-1} \vec{V}] a^{\dagger} \sigma_A^{-} \nonumber \\
+ &[\vec{V}^{\rm T} {\bf M}^{-1} \vec{G}] \sigma_A^{+} a + [\vec{V}^{\rm T} {\bf M}^{-1} \vec{V}] \sigma_A^{+} \sigma_A^{-} \Big) v(t) + \textrm{h.c.} \nonumber  \\
-2 \Big( &[\vec{\gamma}^{\rm T} {\bf M}^{-1} \vec{G}] a v(t) \sigma_A^{+} + [\vec{\gamma}^{\rm T} {\bf M}^{-1} \vec{V}] \sigma_A^{-} v(t) \sigma_A^{+} \Big) + \textrm{h.c.}.
\nonumber
\end{align} 

The third term of Eq.~(\ref{fina_equ_ad}) is calculated similarly, first using $\rho_{gg} (t-\tau-\tau') \approx \rho_{gg} (t-\tau)$ together with $\exp \big[Q \mathcal{L}_0 \tau' \big]\approx \exp \big[ Q \mathcal{L}_B \tau' \big]$, and then $\rho_{gg} (t-\tau) \approx \rho_{gg} (t)$ with $\exp \big[ Q \mathcal{L}_0 \tau \big] \approx \exp \big[ Q \mathcal{L}_B \tau \big]$. Moreover, we use 
\begin{align*}
\mathcal{L}_B  |x_j \rangle \langle x_k^* |  
= \left(\lambda_j + \lambda_k^*\right)  |x_j \rangle \langle x_k^* |,
\end{align*}
with $\langle x_k^* | = \langle g | \big(\vec{x}_k^{*T} \vec{\sigma}_{-}\big)$ and exploit the relations $2 \vec{x}^{\rm T}_{j} \bm{\gamma} \vec{x}^{*}_{\ell}=-(\lambda_{j} + \lambda^{*}_{\ell}) \vec{x}^{\rm T}_{j} \vec{x}^{*}_{\ell}$ and $\vec{X}^{\rm T} {\bf M}^{-1} \vec{V}^{*}=\vec{X}^{\rm T} {\bf M}^{-1} \vec{V}+2 \mi \vec{X}^{\rm T} {\bf M}^{-1} \vec{\gamma}$ with $\vec{X}=\vec{G},\vec{V}$. In the limit of large $t-t_0$ (more precisely averaging over a coarse grained time scale $\Delta t$), we finally obtain  
\begin{align}
&P J \int_{0}^{t-t_0} \hspace{-0.4cm} d \tau e^{Q \mathcal{L}_0 \tau} Q \mathcal{L}_{\rm int} 
\int_{0}^{t-t_{0}-\tau} \hspace{-0.8cm} d \tau' e^{Q \mathcal{L}_0 \tau'} Q \mathcal{L}_{\rm int} v(t-\tau-\tau')  \nonumber \\
\approx & -\mi \Big( [\vec{G}^{\rm T} {\bf M}^{-1} \vec{G}] a v(t) a^{\dagger} + [\vec{G}^{\rm T} {\bf M}^{-1} \vec{V}] a v(t) \sigma_A^{+} \label{third_termm} \\
&+ [\vec{V}^{\rm T} {\bf M}^{-1} \vec{G}] \sigma_A^- v(t) a^{\dagger} + [\vec{V}^{\rm T} {\bf M}^{-1} \vec{V}] \sigma_A^- v(t) \sigma_A^+ \Big) + \textrm{h.c.} \nonumber \\  
&+ 2 \left( [\vec{G}^{\rm T} {\bf M}^{-1} \vec{\gamma}] a v(t) \sigma_A^{+} + [\vec{V}^{\rm T} {\bf M}^{-1} \vec{\gamma}] \sigma_A^{-} v(t) \sigma_A^{+} \right) + \textrm{h.c.}
\nonumber 
\end{align}

Now that the second and third term in the right-hand side of Eq.~(\ref{fina_equ_ad}) have been calculated [Eqs.~(\ref{secon_termm}) and (\ref{third_termm}), respectively], we can now gather these contributions to write Eq.~(\ref{fina_equ_ad}) in the usual master equation form with new effective parameters.  

\subsubsection{Effective master equation parameters}
\label{sec:eff_para}

 Using the property $\vec{X}^{\rm T}{\bf M}^{-1}\vec{Y}=\vec{Y}^{\rm T}{\bf M}^{-1}\vec{X}$ ($\vec{X},\vec{Y}=\vec{G},\vec{V}$) for the symmetric matrix ${\bf M}$, the effective master equation reads
\begin{align}
{\partial_t} v = \mathfrak{L}^{\text{eff}}v = - \mi[H_0^{\text{eff}} + H_{\rm JC}^{\rm eff}, v] + \mathcal{L}^{\text{eff}}v
\label{eq:eff_meq}
\end{align}
with the effective Hamiltonians
\begin{align*}
H_0^{\text{eff}} &= \Delta^{\text{eff}}_{A} \sigma^{+}_{A}\sigma^{-}_{A} + \Delta^{\text{eff}}_{c} a^{\dagger} a \\
H_{\rm JC}^{\rm eff} &= g_A^{\rm eff} \Big( a^{\dagger} \sigma_A^- + \sigma_A^+ a \Big),
\end{align*}
and the effective dissipator
\begin{align}
\label{eq:L^eff}
\mathcal{L}^{\text{eff}} v =
&-\kappa^{\text{eff}} \mathcal{D}(a^{\dagger}, a) v -\gamma_A^{\text{eff}}  \mathcal{D}(\sigma_A^+, \sigma_A^-) v \\ 
&- \mu \Big( \mathcal{D} (a^{\dagger}, \sigma_A^-) v + \mathcal{D} (\sigma_A^+, a) v \Big). \nonumber
\end{align}
Here, the effective parameters are 
\begin{align}
\Delta_c^{\text{eff}} &= \Delta_c - \Re [\vec{G}^{\rm T} {\bf M}^{-1} \vec{G}]
& 
\Delta_A^{\text{eff}} &= - \Re [\vec{V}^{\rm T} {\bf M}^{-1} \vec{V}]
\nonumber \\ 
g_A^{\text{eff}} &= g_A - \Re [\vec{G}^{\rm T} {\bf M}^{-1} \vec{V}] %
&
\kappa^{\text{eff}} &= \kappa + \Im [\vec{G}^{\rm T} {\bf M}^{-1} \vec{G}] \nonumber \\
\gamma_A^{\text{eff}} &= \gamma_A + \Im [\vec{V}^{\rm T} {\bf M}^{-1} \vec{V}]  
&
\mu &= \Im [\vec{G}^{\rm T} {\bf M}^{-1} \vec{V}].
\label{eq:meq_parameters}
\end{align}
Here, dissipative couplings between $A$ and $B$ (and thus $\mathcal{S}$ and $B$) are encoded in $\vec{\gamma} = - \Im[\vec{V}]$.
  
Note that since the emitters of the eliminated ensemble $B$ are supposed to remain close to their ground states (low spin excitations), the same derivation can be carried out when $B$ consists of bosonic degrees of freedom instead of spins. 
 
The effective master equation parameters in Eq.~(\ref{eq:meq_parameters}) are the main result of the adiabatic elimination procedure. In Sec.~\ref{sec:discussion} below, we will analyze those parameters for various situations.

\subsection{Elimination in the classical limit}
\label{sec:class_elim}
 
We now show that the parameters from Eq.~(\ref{eq:meq_parameters}) are identical to parameters that appear in a linear classical model for the limit of low excitation numbers.

The equations of motion for the expectation values of the photon annihilation operator $a$ and the spin lowering operators $\sigma_A^-$ and $\sigma_j^-$ are derived from Eq.~(\ref{full_qme}) as~\cite{Carmichael1989}
\begin{align*}
\partial_t \langle a \rangle = -&\mi \left(\Delta_c - \mi \kappa\right)  \langle a \rangle -\mi g_A \langle \sigma_A^- \rangle -\mi \sum_{j} g_j \langle \sigma_j^- \rangle, \\
\partial_t \langle \sigma_A^- \rangle =-&\gamma_{A} \langle \sigma_A^- \rangle - \mi g_{A} \langle a \rangle - \sum_{j} \left(\gamma_{jA} + \mi \Omega_{j A} \right)\langle \sigma_j^- \rangle, \\
\partial_t \langle \sigma_j^- \rangle 
=-&\mi \left(\Delta_{B}- \mi\gamma_{B}\right)  \langle \sigma_j^- \rangle - \mi g_{j} \langle a \rangle \\
- & \left(\gamma_{jA} + \mi \Omega_{j A} \right) \langle \sigma_A^- \rangle - \sum_{\ell \neq j} \left(\gamma_{j\ell} + \mi \Omega_{j \ell} \right) \langle \sigma_{\ell}^- \rangle,
\end{align*}
under the assumption of low spin excitations $[\sigma_A^-, \sigma_A^+] \approx 1$ and $[\sigma_{\ell}^-, \sigma_{\ell}^+] \approx 1$~\cite{James1993}. Note that since these equations of motion describe coupled harmonic oscillators, this method can alternatively be used when considering another cavity or a mechanical oscillator for $B$ and/or $A$ (see for instance Refs.~\cite{Liu_Cohere_2014,Ramirez-Munoz_Indire_2018}).  

It is convenient to introduce the notations $\alpha \equiv \langle a \rangle$, $\beta_A \equiv \langle \sigma_A^- \rangle$, and $\vec{\beta} \equiv \langle \vec{\sigma}_{-} \rangle$, which allows to write the previous set of equations in the compact form 
\begin{align}
\partial_t \alpha &= - \mi \big[ \Delta_c - \mi \kappa \big] \alpha - \mi g_A \beta_A - \mi \vec{G}^{\rm T} \vec{\beta},  \label{Eq1_sys}\\
\partial_t \beta_{A} &= - \mi \big[ - \mi \gamma_A \big] \beta_A - \mi g_A  \alpha - \mi \vec{V}^{\rm T} \vec{\beta},  \label{Eq2_sys} \\
\partial_t \vec{\beta} &= -\mi {\bf M} \vec{\beta} - \mi \vec{G} \alpha -\mi \vec{V} \beta_A, 
\label{sys}    
\end{align}
where $\vec{G}$ and $\vec{V}$ are defined in Sec.~\ref{sec:decomp_feca}. Equation (\ref{sys}) admits the steady-state solution
\begin{align}
\vec{\beta} = - {\bf M}^{-1} \vec{G} \alpha - {\bf M}^{-1} \vec{V} \beta_{A}.
\label{st_st_sol}
\end{align}

Now using the solution Eq.~(\ref{st_st_sol}) in Eqs.~(\ref{Eq1_sys}) and (\ref{Eq2_sys}) one obtains  
\begin{align}
\partial_t \alpha &= - \mi \big[ \Delta_c^{\text{eff}} - \mi \kappa^{\text{eff}} \big] \alpha - \mi \big[ g_A^{\text{eff}} - \mi \mu \big] \beta_A \nonumber \\
\partial_t \beta_A &= 
- \mi \big[ \Delta_A^{\text{eff}} - \mi \gamma_A^{\text{eff}} \big] \beta_A 
- \mi \big[ g_A^{\text{eff}} - \mi \mu \big]  \alpha,
\label{sys_formal}  
\end{align}
with the same effective parameters as in Eq.~(\ref{eq:meq_parameters}). Equation (\ref{sys_formal}) describes the effective dynamics for the expectation values of the photon and spin operators $a$ and $\sigma^{-}_{A}$, respectively, after adiabatic elimination of the degrees of freedom of $B$.  In Appendix~\ref{sec:validity_class_elim} we show that this procedure is generally valid for a situation where the full adiabatic elimination discussed in Sec.~\ref{sec:full_elim} is valid, i.e.~when the eigenvalues of the matrix ${\bf M}$ (or equivalently ${\bf B}$) associated to the internal dynamics of $B$ are the largest parameters of the problem.

We have thus shown that the time evolution of the subsystem $\mathcal{S}$ in the classical limit [Eq.~\eqref{sys_formal}] corresponds to that of the classical fields $\alpha$ and $\beta_{A}$ with oscillation frequencies $\Delta_c^{\text{eff}}$ and $\Delta_A^{\text{eff}}$, decay rates $\kappa^{\text{eff}}$ and $\gamma_A^{\text{eff}}$, respectively, as well as a coupling $\propto g_A^{\text{eff}} - \mi \mu$ between them.   
  
\section{Discussion of effective parameters}
\label{sec:discussion}

In this section we analyze how the physical parameters of the subsystem $\mathcal{S}$ [Eq.~\eqref{eq:meq_parameters}] are modified, and what their limits are, depending on the system parameters and the geometry. Therefore, we consider the case where $B$ is reduced to a single emitter and provide analytical formulas for the effective parameters in this case [Sec.~\ref{sec:eff_singlespin}]. We find that the presence of $B$ results in a modification of the cavity coupling strength of $A$ (Sec.~\ref{sec:single_em_coup_str}), a change of the linewidths (Sec.~\ref{sec:single_em_eff_linewu}), and leads to joint dissipative processes (Sec.~\ref{sec:single_em_off-diagrate}). In the second part of this section [Sec.~\ref{sec:coll_diss}] we analyze the consequences of the joint dissipative processes on the system dynamics. We discuss modifications of cavity transmission spectra [Sec.~\ref{sec:trans_spec}] and the modification of the on-set of strong cavity-coupling of $A$ [Sec.~\ref{sec:onset_sc}].

\subsection{Effective parameters for \texorpdfstring{$B$}{} being a single emitter}
\label{sec:eff_singlespin}

\begin{figure*}[t]
	\centering
	\includegraphics[width=0.8\textwidth]{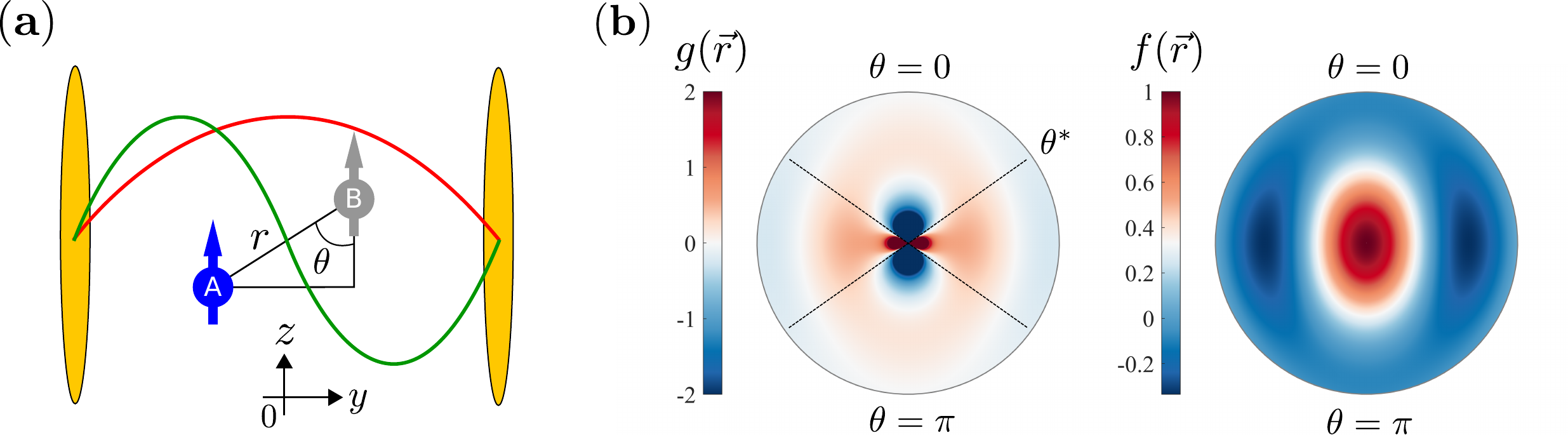}
	\caption{({\bf a}) Two dipoles $A$ and $B$ aligned in the $z$ direction and separated by the vector $\vec{r}=(r,\theta)$ interact via dipole-dipole couplings and with a cavity mode. The first two modes of the cavity are sketched as red and green lines. ({\bf b}) The coherent [$g(\vec{r})$] and dissipative [$f(\vec{r})$] dipole-dipole couplings are plotted for polar coordinates $(r,\theta)$. The outer circles correspond to $r/\lambda=1$ ($\lambda$: cavity mode wavelength). Since the function $g(\vec{r})$ diverges for $r\to 0$, it is plotted for $g(\vec{r}) \in [-2,2]$. The thin dashed lines in the $g(\vec{r})$ profile correspond to the magic angles $\theta^{*}=\arccos (1/\sqrt{3})$ and $\pi-\theta^{*}$ (see text).}
	\label{fig:2}
\end{figure*}

We first consider the situation where $B$ can be considered as a single emitter at position $\vec{r}_{B}\equiv (x_{B},y_{B},z_{B})$ separated from $A$ by $\vec{r} \equiv (x,y,z) \equiv \vec{r}_{B}-\vec{r}_{A}$ with $\vec{r}_{A}\equiv (x_{A},y_{A},z_{A})$ [see Fig.~\ref{fig:2}({\bf a})]. The coupling strength between the $A$ [$B$] and the cavity mode reads $g_{A}=g^{(0)}_{A}\cos\left(k y_{A}\right)$ [$g_{B}=g^{(0)}_{B}\cos\left(k y_{B}\right)$]. Here, $g^{(0)}_{A},g^{(0)}_{B}>0$ and $k=2\pi/\lambda=\omega_{c}/c$ denotes the cavity-photon wave vector (along the $y$-direction) with $\lambda$ the cavity-mode wavelength and $c$ the speed of light in vacuum. 

The dipole-dipole interaction strength between $A$ and $B$ is~\cite{Lehmberg1970_Radia} 
\begin{align}
V_{AB} &= - \frac{3 \sqrt{\gamma_A \gamma_B}}{2} \Big( \sin^2(\theta) \frac{\exp(\mi \xi)}{\xi} \nonumber \\
&+ \Big[ 3 \cos^2(\theta) - 1 \Big] \Big[ \frac{\exp(\mi \xi)}{\xi^{3}} - \mi \frac{\exp(\mi \xi)}{\xi^{2}} \Big] \Big), 
\label{eq:VAB}
\end{align}
with $\xi = k r$ ($r\equiv |\vec{r}|$) and $\theta = \arccos\left( z/r \right)$. Using $\vec{V}\equiv V_{AB}=\Omega_{AB} - \mi \gamma_{AB}$, one can compute the quantities $\vec{G}^{\rm T} {\bf M}^{-1} \vec{G}$, $\vec{G}^{\rm T} {\bf M}^{-1} \vec{V}$, and $\vec{V}^{\rm T} {\bf M}^{-1} \vec{V}$ entering Eq.~(\ref{eq:meq_parameters}), and the effective parameters for the evolution according to Eqs.~\eqref{eq:eff_meq} and \eqref{eq:L^eff} become
\begin{align}
g_A^{\text{eff}} &= g_A - g_B \frac{\Omega_{AB} \Delta_B + \gamma_{AB} \gamma_B }{\Delta_B^2 + \gamma_B^2} \label{eq:eff_gA} \\
\gamma_A^{\text{eff}} &= \gamma_A + \frac{\gamma_B (\Omega_{AB}^2 - \gamma_{AB}^2) - 2 \Delta_B \Omega_{AB} \gamma_{AB}}{\Delta_B^2 + \gamma_B^2} \label{eq:eff_gamA} \\
\kappa^{\text{eff}} &=\kappa + \frac{g^{2}_{B} \gamma_B}{\Delta_B^2 + \gamma_B^2} \label{eq:eff_kappa} \\
\mu &= g_B \frac{\gamma_{B} \Omega_{AB} - \Delta_B \gamma_{AB}}{\Delta_B^2 + \gamma_B^2}
\label{eq:eff_mu}
\end{align}
and
\begin{align*}
\Delta_A^{\text{eff}} & = -\frac{(\Omega_{AB}^2 - \gamma_{AB}^2) \Delta_B + 2 \Omega_{AB} \gamma_{AB} \gamma_B}{\Delta_B^2 + \gamma_B^2} \\
\Delta_c^{\text{eff}} & = \Delta_c - \frac{g^{2}_{B} \Delta_B}{\Delta_B^2 + \gamma_B^2}. 
\end{align*}
The two parameters $\Delta_B$ and $\gamma_B$ are limited by the condition of the validity of the adiabatic elimination, i.e.~the effective model. For instance, they have to fulfill:
\begin{align}
    \max(|\Omega_{AB}|, |\gamma_{AB}|) \ll |\Delta_B - \mi \gamma_B|
    \label{eq:adb_validity}
\end{align}

In the following discussion, we use the two dimensionless functions $g (\vec{r})$ and $f (\vec{r})$ corresponding to the coherent and dissipative parts of the dipole-dipole interaction, 
\begin{align*}
    \Omega_{A B}&= \sqrt{\gamma_A \gamma_B} g (\vec{r})\\
    \gamma_{A B} &= \sqrt{\gamma_A \gamma_B} f (\vec{r})
\end{align*} 
respectively. Their geometry dependence is plotted in Fig.~\ref{fig:2}({\bf b}).

\subsubsection{Effective coupling strength}
\label{sec:single_em_coup_str}

We first focus on the modification of the coupling strength between $A$ and the cavity mode. Using Eq.~(\ref{eq:eff_gA}), the modification of coupling strength, $\Delta g_A \equiv g_A^{\text{eff}} - g_A$ is
\begin{align}
\Delta g_A =  - g_B \Bigg( &\frac{\Omega_{AB}}{|\Delta_B - \mi \gamma_B|} \frac{\Delta_B}{|\Delta_B - \mi \gamma_B|} \nonumber \\
+ &\frac{\gamma_{AB}}{|\Delta_B - \mi \gamma_B|} \frac{\gamma_B}{|\Delta_B - \mi \gamma_B|} \Bigg).
\label{eq:ga_mod}
\end{align}
Therefore, from Eq.~\eqref{eq:ga_mod} and from Eq.~\eqref{eq:adb_validity} it follows that $|\Delta g_A| \ll |g_B^{(0)}|$, which means that $\Delta g_A$ can generally not overcome the cavity coupling strength of $B$, as long as the adiabatic elimination condition remains valid.

Using the position dependence of the cavity couplings, the functions $g(\vec{r})$ and $f(\vec{r})$, and the fact that $g_A^{(0)}/g_B^{(0)}=\sqrt{\gamma_A/\gamma_B}$ we can re-write Eq.~\eqref{eq:ga_mod} as
\begin{align}
{\Delta  g_A}/{g_A^{(0)}} =  - \cos(k y_B) \frac{\gamma_B}{\Delta_B^2 + \gamma_B^2} \Bigg( \Delta_B g(\vec{r}) + \gamma_B f(\vec{r}) \Bigg).
\label{eq:ga_mod_coh_vs_diss}
\end{align}
Evidently, the change of the cavity coupling of $A$ can be either induced by coherent [$\propto \Delta_B g(\vec{r})$] or by dissipative [$\propto \gamma_B f(\vec{r})$] dipole-dipole interactions in Eq.~\eqref{eq:ga_mod_coh_vs_diss}. Let us now analyze those two limits in more detail:

The dissipative limit $|\gamma_B f(\vec{r})| \gg |\Delta_B g(\vec{r})|$ is difficult to achieve in the near-field, purely by the geometry. This is due to the divergence $|g(\vec{r})| \propto 1/r^3$ while $|f(\vec{r})| \to 1$ for $r \ll \lambda$. An exception are ``magic angle'' dipole configurations with $3\cos^2(\theta) = 1$, for which near-field terms vanish in Eq.~\eqref{eq:VAB}, then only implying $|g(\vec{r})| \propto 1/r$ for $r \to 0$ [see Fig.~\ref{fig:2}({\bf b})]. In general, however, the dissipative limit is achieved for $\Delta_B \approx 0$, i.e.~if $B$ is close to resonance with $A$ and the cavity. In this case the validity of the adiabatic elimination in Eq.~\eqref{eq:adb_validity} must be ensured by large $\gamma_B$, i.e.~by $\gamma_B \gg \gamma_A g^2(\vec{r})$ and $\gamma_B \gg\gamma_A f^2(\vec{r})$. The modification of the cavity coupling becomes
\begin{align*}
{\Delta  g_A}/{g_A^{(0)}}=-\cos(k y_B) f(\vec{r}).
\end{align*}
Since $|f(\vec{r})| < 1$, it follows that $|\Delta g_A/g_A^{(0)}| < 1$ and the modification of the coupling strength is therefore generally limited by $g_A^{(0)}$. Still, coherent coupling of $A$ to the cavity can be effectively induced, for instance, when $A$ is at a node of the cavity mode ($g_A = 0$) and $B$ in the near-field of $A$ with $r \ll \lambda$ and $\cos(k y_B) \neq 0$~\cite{Schutz_Ensemb_2019}.

The coherent limit $|\Delta_B g(\vec{r})| \gg |\gamma_B f(\vec{r})|$ is generally achieved for most configurations in the near-field where $|g(\vec{r})| \gg |f(\vec{r})|$, as long as $|\Delta_B| \gtrsim \gamma_B$. The modification of the coupling,
\begin{align*}
{\Delta  g_A}/{g_A^{(0)}}=-\cos(k y_B) g(\vec{r}) \frac{\Delta_B \gamma_B}{\Delta_B^2 + \gamma_B^2},
\end{align*}
is then generally limited by the magnitude of $g(\vec{r})$. In a regime where the adiabatic elimination condition~\eqref{eq:adb_validity} is valid, if $|\Delta_B| \sim \gamma_B$ it is required that $|\Omega_{AB}| \ll \gamma_B$ and thus $|g(\vec{r})| \ll \sqrt{\gamma_{B}/\gamma_{A}}$. On the other hand, the condition~\eqref{eq:adb_validity} can be easily fulfilled for large detuning, $|\Delta_B| \gg \gamma_B$, in this case
\begin{align*}
{\Delta  g_A}/{g_A^{(0)}}=-\cos(k y_B) g(\vec{r}) \frac{\gamma_B}{\Delta_B} \ll |g(\vec{r})|.
\end{align*}
While the change in coupling strength is still limited by $|g(\vec{r})|$, generally large enhancements of $g_A$ are possible in this regime \cite{Schutz_Ensemb_2019}.

\subsubsection{Effective linewidths}
\label{sec:single_em_eff_linewu}

 We now focus on the linewidth modifications of the cavity and $A$. From Eq.~(\ref{eq:eff_kappa}), it is evident that the cavity mode linewidth $\kappa$ can only be broadened by the interaction with $B$ since 
\begin{align*}
     \Delta\kappa \equiv \kappa^{\text{eff}} - \kappa =\frac{ g_B^2 \gamma_B }{\Delta_B^2 + \gamma_B^2} \geq 0.
\end{align*}
The change in effective linewidth of $A$, $\Delta \gamma_A \equiv \gamma_A^{\text{eff}} - \gamma_A$, using Eq.~\eqref{eq:eff_gamA}, can be written in the form
\begin{align}
\Delta \gamma_A/\gamma_A = \bigg[  \frac{[\gamma_B g(\vec{r}) - \Delta_B f(\vec{r})]^{2}}{\Delta_B^2 + \gamma_B^2} - f^2(\vec{r})  \bigg].
\label{eq:eff_gamA_mod}
\end{align}
The first term in Eq.~\eqref{eq:eff_gamA_mod},
\begin{align}
\frac{[\gamma_B g(\vec{r}) - \Delta_B f(\vec{r}) ]^2}{\Delta_B^2 + \gamma_B^2} \geq 0,
\label{eq:gam_broad}
\end{align}
always leads to a broadening of $\gamma_A$, while the second term $f^2(\vec{r}) \geq 0$ always contributes to a linewidth reduction. 

Importantly, strong linewidth narrowing is possible for special points for which the broadening from Eq.~\eqref{eq:gam_broad} vanishes, i.e.~when
\begin{align}
\gamma_B g(\vec{r}) = \Delta_B f(\vec{r})
\label{eq:ideal_gamA_narrow}
\end{align}
is fulfilled. Then, one can achieve  large linewidth reductions $\Delta \gamma_A/\gamma_A \to -1$ in the near field, where $f(\vec{r})\to 1$. Furthermore, we note that a reduction can also be achieved in the dissipative limit with $\Delta_B =0$, discussed above. Then, 
\begin{align*}
\Delta \gamma_A/ \gamma_A =  \left[g^2(\vec{r}) - f^2(\vec{r})\right],
\end{align*}
and $\gamma_A^{\text{eff}}<\gamma_A$ occurs for any geometry with $|f(\vec{r})| > |g(\vec{r})|$, e.g., close to the magic angle $\theta^{*}=\arccos (1/\sqrt{3})$ in the near-field [see Fig.~\ref{fig:2}\textbf{(b)}].

\subsubsection{Joint dissipative couplings}
\label{sec:single_em_off-diagrate}

We now focus on the effective parameter $\mu$ that enters as collective dissipative term in Eq.~\eqref{eq:L^eff}. Defining the (positive) broadening of $A$ as
\begin{align*}
\delta \gamma_A &\equiv \gamma_A \frac{[\gamma_B g(\vec{r}) - \Delta_B f(\vec{r}) ]^2}{\Delta_B^2 + \gamma_B^2}, 
\end{align*}
the modulus of $\mu$ [Eq.~(\ref{eq:eff_mu})] can be written in the very simple form 
\begin{align*}
|\mu| = \sqrt{\Delta \kappa \delta \gamma_A}.
\end{align*}
Since $\delta \gamma_A < \gamma_A^{\text{eff}}$ and $\Delta \kappa < \kappa^{\text{eff}}$, this implies that $|\mu| < \sqrt{\kappa^{\text{eff}} \gamma^{\text{eff}}_A} \leq \text{max} (\kappa^{\text{eff}}, \gamma_A^{\text{eff}})$, which sets a fundamental limitation on $|\mu|$. Note that at special points of ideal linewidth narrowing [condition~\eqref{eq:ideal_gamA_narrow}] $\delta \gamma_A =0$ and then also $\mu = 0$.

In order to gain further insights on the role of $\mu$, it is useful to diagonalize the dissipator introduced in Eq.~\eqref{eq:L^eff}. By defining Lindblad operators
\begin{align*}
L_+ &= \cos(\alpha/2) a  + \sin(\alpha/2) \sigma_A^{-}, \\
L_- &= -\sin(\alpha/2) a  + \cos(\alpha/2) \sigma_A^{-},
\end{align*}
Eq.~\eqref{eq:L^eff} can be written as
\begin{align*}
\mathcal{L}^{\text{eff}} v 
= 
- \gamma_+
\mathcal{D}(L^\dag_+, L_+) v
- \gamma_-
\mathcal{D}(L^\dag_-, L_-) v.
\end{align*}

The Lindblad operators are linear combinations of the photon annihilation operator $a$ and the spin lowering operator 
$\sigma_A^-$, with $\tan(\alpha) = 2 \mu /(\kappa^{\rm eff} - \gamma^{\rm eff}_A)$ and $0\leq \alpha < 2 \pi$. 
The decay rates associated to the Lindblad operators are
\begin{align*}
\gamma_\pm = \frac{ \kappa^{\rm eff}  + \gamma_A^{\rm eff}}{2} \pm \sqrt{\frac{ (\kappa^{\rm eff} - \gamma_A^{\rm eff})^2}{4} + \mu^2},
\end{align*}
and correspond to the eigenvalues of the matrix 
\begin{align*}
\begin{pmatrix}
\kappa^{\rm eff} & \mu \\
\mu              & \gamma_A^{\rm eff} \\
\end{pmatrix}.
\end{align*}
The two Lindblad jump operators describe mutual decay processes between the cavity and $A$, which are mediated by the presence of $B$. 
This is analogous to sub- and superradiant decay of atoms due to collective incoherent processes, induced by the coupling to a joint cavity mode~\cite{Bonifacio1971,Shammah2018}. 
Similarly, also the parameters $\gamma_{j A}$ introduced in Sec.~\ref{sec:full_qme} correspond to such off-diagonal decay mechanisms, which are in this case mediated by the surrounding electromagnetic field~\cite{Lehmberg1970_Radia}.
 
\subsection{Consequences of joint dissipative coupling}
\label{sec:coll_diss}

Here, we investigate the consequences of the mutual decay mechanisms between the cavity mode and $A$ in the effective model that is mediated by the presence of the ensemble $B$. We analyze the modification of the cavity transmission spectum in Sec.~\ref{sec:trans_spec} and study the modification of the on-set of strong coupling between the cavity and $A$ in Sec.~\ref{sec:onset_sc}.

\subsubsection{Cavity transmission spectrum}
\label{sec:trans_spec}

To compute a cavity transmission spectrum, we consider a weak laser probe driving the cavity, described by the (additional) time-dependent Hamiltonian 
\begin{align*}
H_{L} = \eta \big( a e^{\mi \omega_L t} + a^{\dagger} e^{-\mi \omega_L t} \big)
\end{align*}
with frequency $\omega_L$ and strength $\eta$.

Similarly as in Sec.~\ref{sec:class_elim}, we derive the equations of motion in the classical linear limit valid for low excitation numbers (initial state without excitations and  weak drive $\eta \to 0$). Then, using the classical variables $\alpha \equiv \langle a \rangle$, $\beta_A \equiv \langle \sigma_A^- \rangle$, and $\vec{\beta} \equiv \langle \vec{\sigma}_- \rangle$, in the frame rotating with the laser frequency $\omega_L$, the equations of motion are
\begin{align*}
\partial_t \alpha &= - \mi \big[ \widetilde{\Delta}_c - \mi \kappa \big] \alpha - \mi g_A \beta_A - \mi \vec{G}^{\rm T} \vec{\beta} - \mi \eta  \\
\partial_t \beta_A &= - \mi \big[ \widetilde{\Delta}_A - \mi \gamma_A \big] \beta_A - \mi g_A \alpha - \mi \vec{V}^{\rm T} \vec{\beta} \\
\partial_t \vec{\beta} &= -\mi \widetilde{{\bf M}} \vec{\beta} - \mi \vec{G} \alpha -\mi \vec{V} \beta_A,
\end{align*}
where $\widetilde{\Delta}_c = \omega_c - \omega_L$, $\widetilde{\Delta}_A = \omega_A - \omega_L$ and  $(\widetilde{{\bf M}})_{j \ell}= \big(\widetilde{\Delta}_{B} - \mi \gamma_B\big) \delta_{j\ell} + (1-\delta_{j\ell})(\Omega_{j\ell} - \mi \gamma_{j \ell})$ with $\widetilde{\Delta}_B = \omega_B - \omega_L$. The steady-state solution is
\begin{align}
\begin{pmatrix}
\alpha^{\rm st} \\ \\
\beta^{\rm st}_A
\end{pmatrix}
=
-
\begin{pmatrix}
\widetilde{\Delta}_{c}^{\text{eff}} - \mi \widetilde{\kappa}^{\text{eff}} 
& \widetilde{g}_A^{\text{eff}} - \mi \widetilde{\mu} \\ \\
\widetilde{g}_A^{\text{eff}} - \mi \widetilde{\mu}
& \widetilde{\Delta}_{A}^{\text{eff}} - \mi \widetilde{\gamma}_A^{\text{eff}}
\end{pmatrix}^{-1}
\begin{pmatrix}
\eta \\ \\
0 
\end{pmatrix}
,
\label{eq:mat_laser}
\end{align}
with the definitions
\begin{align*}
\widetilde{\Delta}_c^{\text{eff}} &= \widetilde{\Delta}_c - \Re [\vec{G}^{\rm T} \widetilde{{\bf M}}^{-1} \vec{G}]
& 
\widetilde{\Delta}_A^{\text{eff}} &= \widetilde{\Delta}_A - \Re [\vec{V}^{\rm T} \widetilde{{\bf M}}^{-1} \vec{V}]
\nonumber \\ 
\widetilde{g}_A^{\text{eff}} &= g_A - \Re [\vec{G}^{\rm T} \widetilde{{\bf M}}^{-1} \vec{V}] &
\widetilde{\kappa}^{\text{eff}} &= \kappa + \Im [\vec{G}^{\rm T} \widetilde{{\bf M}}^{-1} \vec{G}] \nonumber \\
\widetilde{\gamma}_A^{\text{eff}} &= \gamma_A + \Im [\vec{V}^{\rm T} \widetilde{{\bf M}}^{-1} \vec{V}]  
&
\widetilde{\mu} &= \Im [\vec{G}^{\rm T} \widetilde{{\bf M}}^{-1} \vec{V}].
\end{align*}

The cavity transmission spectrum is proportional to the mean photon number in the steady-state $\propto |\alpha^{\rm st}|^2$, which can be obtained from Eq.~\eqref{eq:mat_laser}.

We consider a situation where the eigenvalues of the matrix in the laser frame, $\widetilde{{\bf M}} = {\bf M} + \id (\omega_A - \omega_L)$ ($\id$ is the identity matrix), are well approximated by the eigenvalues of $\bf{M}$. This is true when the shift $|\omega_A - \omega_L|$ is small compared to the real part of the eigenvalues of $\bf{M}$, which are related to the eigenfrequencies of the interacting $B$ ensemble. Note that this condition can be ensured in the dispersive limit (where the eigenfrequencies of $B$ are far-detuned, i.e.~spectrally well separated from $\mathcal{S}$) and that this is consistent with the requirement for the validity of the adiabatic elimination (see Appendix~\ref{sec:validity_class_elim}), also in the presence of the laser-drive. The probe laser is then scanned only over a frequency range that is relevant for the dynamics of the subsystem $\mathcal{S}$.  In this situation $\widetilde{\kappa}^{\text{eff}} \approx \kappa^{\text{eff}}$, $\widetilde{\gamma}_A^{\text{eff}} \approx \gamma_A^{\text{eff}}$, $\widetilde{g}_A^{\text{eff}} \approx g_A^{\text{eff}}$, $\widetilde{\mu} \approx \mu$, and the laser frequency only enters through the detunings $\widetilde{\Delta}_c^{\text{eff}} = \omega_c^{\rm eff} - \omega_L$ and $\widetilde{\Delta}_A^{\text{eff}} = \omega_A^{\rm eff} - \omega_L$ with $\omega_c^{\rm eff} \simeq \omega_c - \Re [\vec{G}^{\rm T} {\bf M}^{-1} \vec{G}]$ and $\omega_A^{\rm eff} \simeq \omega_A - \Re [\vec{V}^{\rm T} {\bf M}^{-1} \vec{V}]$. Then, we define the normalized steady-state cavity transmission spectrum by $\mathcal{T}_{c} (\omega_{L}) \equiv (\kappa^2/\eta^2) |\alpha^{\rm st}|^2$ with
\begin{align*}
\alpha^{\rm st} = \frac{\eta\left(\widetilde{\Delta}_A^{\text{eff}} -\mi \gamma_A^{\text{eff}}\right)}{\left(g_A^{\text{eff}}- \mi \mu \right)^{2} - \left(\widetilde{\Delta}_{A}^{\text{eff}} - \mi \gamma_A^{\text{eff}} \right) \left(\widetilde{\Delta}_{c}^{\text{eff}} -\mi  \kappa^{\text{eff}} \right)}.
\end{align*}

\begin{figure}[t]
\centering
\includegraphics[width=1\columnwidth]{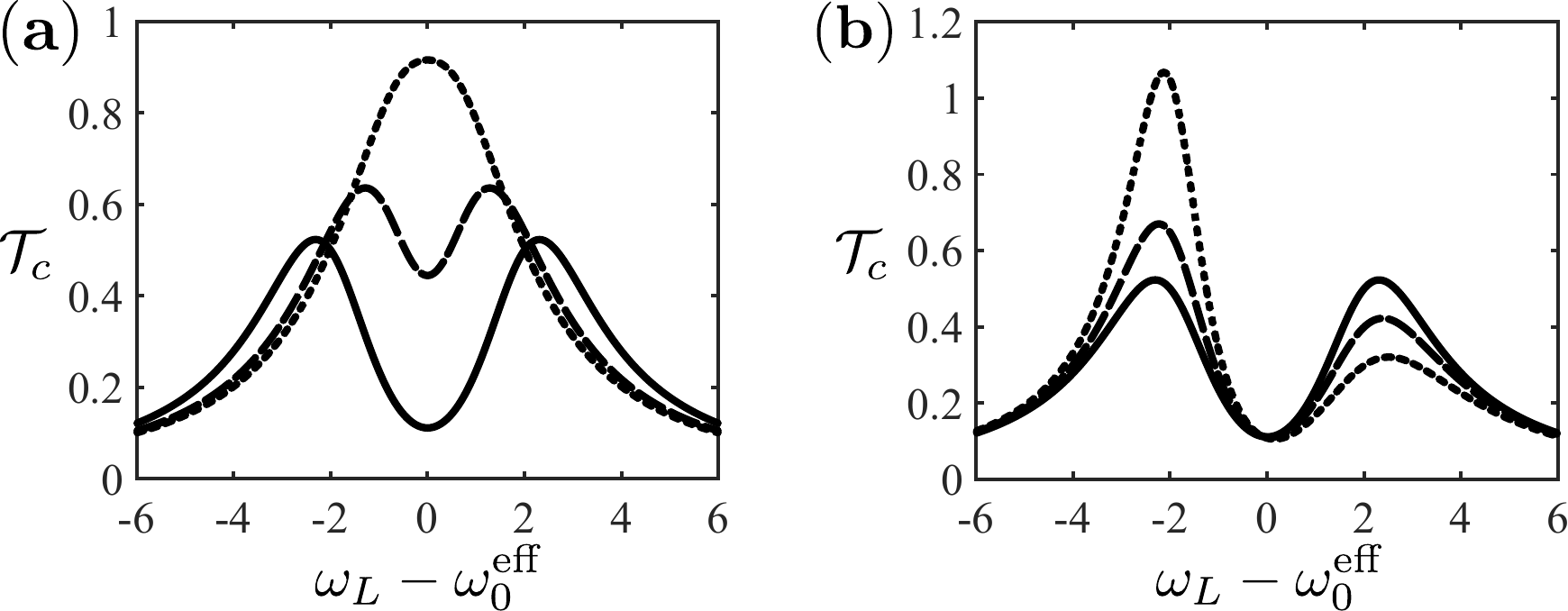}
\caption{Normalized steady-state cavity transmission spectra $\mathcal{T}_{c}$. (\textbf{a}) $\mu=0$: Dotted, dashed, and solid lines correspond to $g^{\text{eff}}_{A}=0.3,1,2$, respectively. (\textbf{b}) $g^{\text{eff}}_{A}=2$: Solid, dashed, and dotted lines correspond to $\mu=0,0.2,0.5$, respectively. Other parameters are $\gamma^{\text{eff}}_A=1$, $\kappa^{\text{eff}}=2$, and $\eta=0.1$.}
\label{fig:cavtra}
\end{figure}

For a finite mutual decay rate $\mu \neq 0$, we find that this transmission spectrum features two asymmetric peaks separated by an energy splitting $\propto 2 |g_A^{\text{eff}}|$ for large enough $g_A^{\text{eff}}$. It is instructive to associate the peaks to ``polaritons'' by diagonalizing the matrix entering Eq.~(\ref{eq:mat_laser}). Considering $A$ in resonance with the cavity mode, i.e.~$\omega_A^{\rm eff} = \omega_c^{\rm eff} \equiv \omega_0^{\rm eff}$, and by defining $\widetilde{\Delta}^{\text{eff}} = \omega_0^{\rm eff} - \omega_L$, Eq.~(\ref{eq:mat_laser}) can be decomposed as (assuming the dispersive limit):
\begin{align*}
\begin{pmatrix}
\alpha^{\rm st} \\ \\
\beta^{\rm st}_A
\end{pmatrix}
&= \Bigg\{
-\mi
{\bf T} 
- \begin{pmatrix}
\widetilde{\Delta}^{\text{eff}} 
& 0 \\ \\
0
& \widetilde{\Delta}^{\text{eff}}
\end{pmatrix}
\Bigg\}^{-1}
\begin{pmatrix}
\eta \\ \\
0
\end{pmatrix},
\end{align*}
where
\begin{align*}
{\bf T} = \begin{pmatrix}
- \kappa^{\text{eff}} 
& -\mi g_A^{\text{eff}} - \mu \\ \\
-\mi g_A^{\text{eff}} - \mu
& - \gamma_A^{\text{eff}}
\end{pmatrix}  \nonumber
\end{align*}
is a non-hermitian, complex symmetric matrix, which we diagonalize as ${\bf T}=\sum_{p=\pm}\xi_{p} \vec{u}_{p}\vec{u}^{\rm T}_{p}$. Here, the two eigenvectors are defined as $\vec{u}_{+}=(u^{+}_{1},u^{+}_{2})$ and $\vec{u}_{-}=(u^{-}_{1},u^{-}_{2})$, with $\sum_{p=\pm} \vec{u}_{p}\vec{u}^{\rm T}_{p}=\id$. The eigenvalues are
\begin{align*}
\xi_{\pm} = -\frac{\gamma^{\text{eff}}_A + \kappa^{\text{eff}}}{2} \mp \sqrt{ \left( \frac{\kappa^{\text{eff}}-\gamma^{\text{eff}}_A}{2} \right)^2 - \left( g^{\text{eff}}_{A} - \mi \mu \right)^2}. 
\end{align*}
We introduce the polariton linewidths and frequencies associated to the real and imaginary parts of $\xi_{\pm}$, namely $\Gamma_{\pm}=-\Re [\xi_{\pm}]$ and $\omega_{\pm} = \omega_0^{\rm eff} - \Im [\xi_{\pm}]$. This leads to
\begin{align}
\mathcal{T}_{c} (\omega_{L})= \kappa^2 \Bigg| \frac{Z_{+}}{\omega_{L}-\omega_{+}+\mi \Gamma_{+}} + \frac{Z_{-}}{\omega_{L}-\omega_{-}+\mi \Gamma_{-}}\Bigg|^{2},
\label{trans_appr}
\end{align}
where $Z_{+}=u^{+}_{1}u^{-}_{2}/(u^{+}_{1}u^{-}_{2}- u^{-}_{1}u^{+}_{2})$ and $Z_{-}=u^{-}_{1}u^{+}_{2}/(u^{-}_{1}u^{+}_{2}- u^{+}_{1}u^{-}_{2})$ are related to the cavity and spin admixtures of the polaritons eigenmodes. 

Cavity transmission spectra from Eq.~(\ref{trans_appr}) are plotted for $\mu = 0$ and different effective coupling strengths $g_A^{\rm eff}$ in Fig.~\ref{fig:cavtra}(\textbf{a}), and for $g_A^{\rm eff} = 2$ and different mutual decay rates $\mu$ in Fig.~\ref{fig:cavtra}(\textbf{b}). For $\mu = 0$, the spectrum is symmetric with respect to $\omega_0^{\rm eff}$, and increasing $g_A^{\rm eff}$ allows to enter the strong coupling regime characterized by the emergence of two well-resolved polariton peaks. We find that increasing $\mu>0$ leads to an asymmetric spectrum with two peaks of different heights and linewidths $\Gamma_+ > \Gamma_-$. In particular, we find that here the joint dissipative processes between $A$ and the cavity lead to a lower/upper polariton with sub-/superradiant linewidth, respectively.

\subsubsection{On-set of strong coupling}
\label{sec:onset_sc}

\begin{figure}[t]
	\centering
	\includegraphics[width=1\columnwidth]{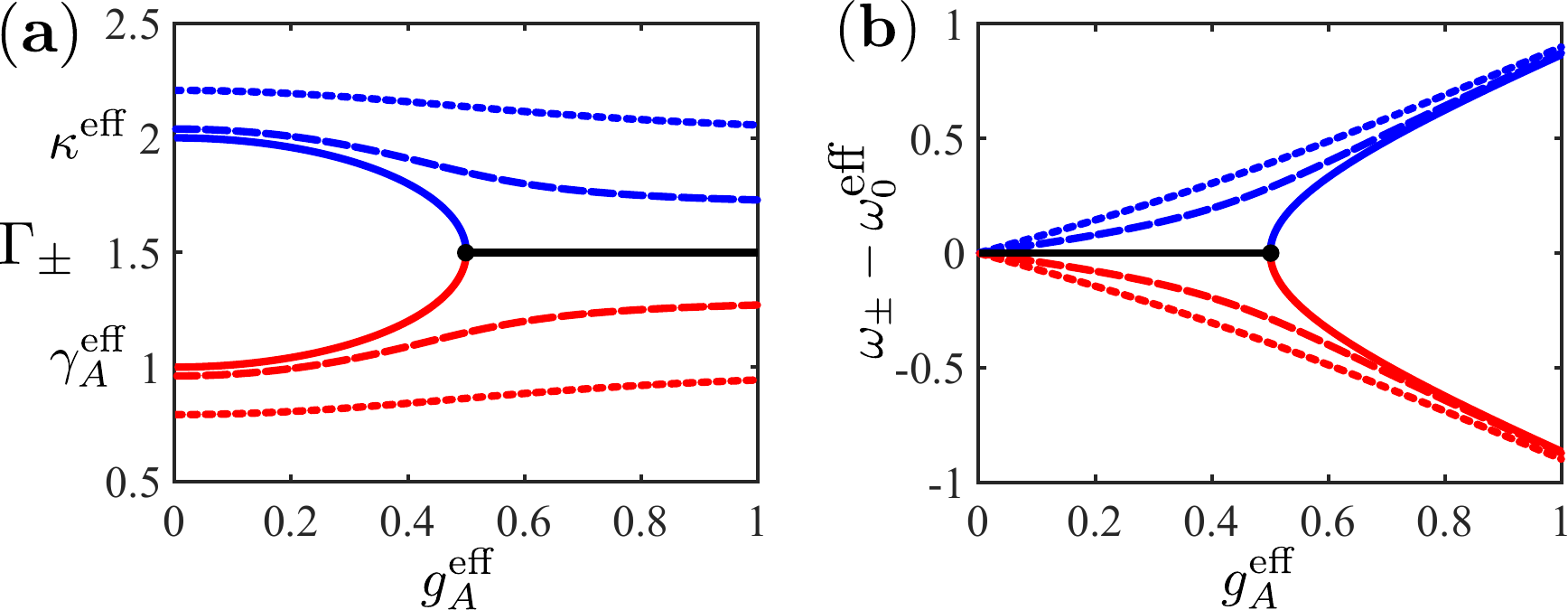}
	\caption{Polariton linewidths $\Gamma_{\pm}$ (\textbf{a}) and frequencies $\omega_{\pm}$ (\textbf{b}) as a function of $g^{\text{eff}}_{A} \geq 0$ for different $\mu \geq 0$. The upper (here $\xi_+$) and lower (here $\xi_-$) polaritons are depicted as blue and red lines, while the solid, dashed, and dotted lines correspond to $\mu=0,0.2,0.5$, respectively. Here, $\gamma^{\text{eff}}_A=1$ and $\kappa^{\text{eff}}=2$.}
	\label{fig:assy}
\end{figure}

Finally, we focus on how a finite $\mu \neq 0$ modifies the conditions for reaching strong coupling between $A$ and the cavity. The polariton linewidths $\Gamma_{\pm}$ and frequencies $\omega_{\pm}$ are plotted in Fig.~\ref{fig:assy}(\textbf{a}) \& (\textbf{b}), respectively, as a function of $g^{\text{eff}}_{A} \geq 0$ for different $\mu \geq 0$. 

We first focus on the case $\mu=0$ (solid lines). For $g^{\text{eff}}_{A} < \vert \kappa^{\text{eff}} - \gamma^{\text{eff}}_A \vert/2$, the two polariton modes are undefined, and the eigenmodes of the system have frequencies $\omega_{\pm}=\omega_0^{\rm eff}$ and linewidths $\Gamma_{\pm}$ with ${\rm min} (\gamma_A^{\rm eff}, \kappa^{\rm eff}) \leq \Gamma_- \leq \Gamma_+ \leq {\rm max} (\gamma_A^{\rm eff}, \kappa^{\rm eff})$. In this weak coupling regime, the cavity transmission spectrum features two strongly overlapping peaks [see Fig.~\ref{fig:cavtra}(\textbf{a})], and the time evolution of the system exhibits overdamped Rabi oscillations. 

For $\mu=0$, the strong coupling is reached when $g^{\text{eff}}_{A} > \vert \kappa^{\text{eff}} - \gamma^{\text{eff}}_A \vert /2$, or alternatively when $g^{\text{eff}}_{A} > \text{max} (\gamma^{\text{eff}}_A,\kappa^{\text{eff}})$. In this case, two polariton modes with different frequencies $\omega_{\pm}$ and identical linewidths, $(\kappa^{\text{eff}} + \gamma^{\text{eff}}_A)/2$, emerge. The transmission spectrum features two weakly overlapping peaks [see Fig.~\ref{fig:cavtra}(\textbf{a})], and the dynamics of $\mathcal{S}$ exhibits well defined Rabi oscillations. The two eigenvalues $\xi_{\pm}$ coalesce for the ``exceptional point'' $g^{\text{eff}}_{A}=(\kappa^{\text{eff}} - \gamma^{\text{eff}}_A)/2$ \cite{Berry2004,Gao2018}.   

Finite $\mu \neq 0$ affects the on-set of strong coupling. In this case, the degeneracy of the two eigenvalues $\xi_{\pm}$ is removed and the exceptional point for the on-set of strong coupling disappears. The polariton linewidths and frequencies are different for all $g^{\text{eff}}_{A}$. By calculating the first derivative of the linewidth $\Gamma_{+}$ with respect to $g^{\text{eff}}_{A} \geq 0$, we find that $\Gamma_{+}$ is a monotonically decreasing function of $g^{\text{eff}}_{A}$. Therefore, the most restrictive condition for strong coupling in the case $\mu\neq 0$ is $g^{\text{eff}}_{A} > \gamma_{+}$, where 
\begin{align*}
\gamma_{+} \equiv \Gamma_{+} (g^{\text{eff}}_{A}=0) = \frac{\gamma^{\text{eff}}_A + \kappa^{\text{eff}}}{2} + \sqrt{ \left( \frac{\kappa^{\text{eff}}-\gamma^{\text{eff}}_A}{2} \right)^2 + \mu^2}. 
\end{align*}

We note that for the discussion of Fig.~\ref{fig:cavtra} and Fig.~\ref{fig:assy} above, $\mu \geq 0$ and $g^{\text{eff}}_{A}\geq 0$ have been assumed. The fact that the lower/upper polariton displays respective sub-/superradiant properties is a consequence of this. The features presented in Figs.~\ref{fig:cavtra} and \ref{fig:assy} hold for arbitrary $\mu$ and $g^{\text{eff}}_{A}$, whereas for $\mu g^{\text{eff}}_{A} < 0$ the asymmetry is reversed, i.e.~the lower (upper) polariton exhibits superradiant (subradiant) behavior.

\section{Conclusion \& Outlook}
\label{sec:conclusion}

 In conclusion, we have carried out a detailed study of the impact of a dipolar environment on the dynamics of a single dipole $A$ coupled to a cavity mode. We performed a detailed adiabatic elimination of the dipoles in the environment and computed effective parameters for the subsystem consisting of $A$ and the cavity. Since the dipoles in the environment couple to both $A$ (with full coherent and dissipative dipole-dipole interactions) and the cavity, they modify the properties of the subsystem significantly. We analyzed effective modifications and limitations of subsystem linewidths, coherent cavity coupling strengths, and emerging collective dissipative processes between $A$ and the cavity. In particular, the latter joint dissipative decay processes can lead to peculiar signatures in the strong coupling regime.
 
The results derived here have been used for the cavity-coupling enhancing scheme proposed in \cite{Schutz_Ensemb_2019} and may prove to be useful for similar schemes using other auxiliary optical or mechanical resonators~\cite{Liu_Cohere_2014,Ramirez-Munoz_Indire_2018}. Enhancement schemes can help reaching the strong coupling regime between a single emitter and a cavity under suitable conditions. Reaching strong coupling of a single emitter may lead to important applications in quantum information processing with photons, due to strong photon non-linearities~\cite{Schutz_Ensemb_2019}.

Furthermore, our work highlights the impact of ``active'' environments on cavity-QED systems, which are relevant in the field of molecular polaritonics~\cite{hutch2012,herrera2016,Flick2017,Ribeiro2018,Thomas2019,cohen2019} where environments such as solvents can play a crucial role~\cite{Lather2018,Thomas_Explor_2019}. There, other interesting perspectives of this work include extensions to disordered ensembles in the context of the emergent field of polaritonic chemistry.  

{\textbf{Acknowledgements} --- }
We are grateful to C.~Genes for stimulating discussions. Work in Strasbourg was supported by the ANR - ``ERA-NET QuantERA'' - Projet ``RouTe'' (ANR-18-QUAN-0005-01) and LabEx NIE. This work is supported by IdEx Unistra (project STEMQuS) with funding managed by the French National Research Agency as part of the ``Investments for the future program''.

\bibliographystyle{apsrev4-2}
\bibliography{main}{}

\begin{thebibliography}{46}%
\makeatletter
\providecommand \@ifxundefined [1]{%
 \@ifx{#1\undefined}
}%
\providecommand \@ifnum [1]{%
 \ifnum #1\expandafter \@firstoftwo
 \else \expandafter \@secondoftwo
 \fi
}%
\providecommand \@ifx [1]{%
 \ifx #1\expandafter \@firstoftwo
 \else \expandafter \@secondoftwo
 \fi
}%
\providecommand \natexlab [1]{#1}%
\providecommand \enquote  [1]{``#1''}%
\providecommand \bibnamefont  [1]{#1}%
\providecommand \bibfnamefont [1]{#1}%
\providecommand \citenamefont [1]{#1}%
\providecommand \href@noop [0]{\@secondoftwo}%
\providecommand \href [0]{\begingroup \@sanitize@url \@href}%
\providecommand \@href[1]{\@@startlink{#1}\@@href}%
\providecommand \@@href[1]{\endgroup#1\@@endlink}%
\providecommand \@sanitize@url [0]{\catcode `\\12\catcode `\$12\catcode
  `\&12\catcode `\#12\catcode `\^12\catcode `\_12\catcode `\%12\relax}%
\providecommand \@@startlink[1]{}%
\providecommand \@@endlink[0]{}%
\providecommand \url  [0]{\begingroup\@sanitize@url \@url }%
\providecommand \@url [1]{\endgroup\@href {#1}{\urlprefix }}%
\providecommand \urlprefix  [0]{URL }%
\providecommand \Eprint [0]{\href }%
\providecommand \doibase [0]{https://doi.org/}%
\providecommand \selectlanguage [0]{\@gobble}%
\providecommand \bibinfo  [0]{\@secondoftwo}%
\providecommand \bibfield  [0]{\@secondoftwo}%
\providecommand \translation [1]{[#1]}%
\providecommand \BibitemOpen [0]{}%
\providecommand \bibitemStop [0]{}%
\providecommand \bibitemNoStop [0]{.\EOS\space}%
\providecommand \EOS [0]{\spacefactor3000\relax}%
\providecommand \BibitemShut  [1]{\csname bibitem#1\endcsname}%
\let\auto@bib@innerbib\@empty
\bibitem [{\citenamefont {Sch{\ifmmode\ddot{u}\else\"{u}\fi}tz}\ \emph
  {et~al.}(2019)\citenamefont {Sch{\ifmmode\ddot{u}\else\"{u}\fi}tz},
  \citenamefont {Schachenmayer}, \citenamefont
  {Hagenm{\ifmmode\ddot{u}\else\"{u}\fi}ller}, \citenamefont {Brennen},
  \citenamefont {Volz}, \citenamefont {Sandoghdar}, \citenamefont {Ebbesen},
  \citenamefont {Genes},\ and\ \citenamefont {Pupillo}}]{Schutz_Ensemb_2019}%
  \BibitemOpen
  \bibfield  {author} {\bibinfo {author} {\bibfnamefont {S.}~\bibnamefont
  {Sch{\ifmmode\ddot{u}\else\"{u}\fi}tz}}, \bibinfo {author} {\bibfnamefont
  {J.}~\bibnamefont {Schachenmayer}}, \bibinfo {author} {\bibfnamefont
  {D.}~\bibnamefont {Hagenm{\ifmmode\ddot{u}\else\"{u}\fi}ller}}, \bibinfo
  {author} {\bibfnamefont {G.~K.}\ \bibnamefont {Brennen}}, \bibinfo {author}
  {\bibfnamefont {T.}~\bibnamefont {Volz}}, \bibinfo {author} {\bibfnamefont
  {V.}~\bibnamefont {Sandoghdar}}, \bibinfo {author} {\bibfnamefont {T.~W.}\
  \bibnamefont {Ebbesen}}, \bibinfo {author} {\bibfnamefont {C.}~\bibnamefont
  {Genes}},\ and\ \bibinfo {author} {\bibfnamefont {G.}~\bibnamefont
  {Pupillo}},\ }\href {https://arxiv.org/abs/1904.08888} {\bibfield  {journal}
  {\bibinfo  {journal} {arXiv}\ } (\bibinfo {year} {2019})},\ \Eprint
  {https://arxiv.org/abs/1904.08888} {1904.08888} \BibitemShut {NoStop}%
\bibitem [{\citenamefont {Pockrand}\ \emph {et~al.}(1982)\citenamefont
  {Pockrand}, \citenamefont {Brillante},\ and\ \citenamefont
  {M\"{o}bius}}]{Pockrand1982}%
  \BibitemOpen
  \bibfield  {author} {\bibinfo {author} {\bibfnamefont {I.}~\bibnamefont
  {Pockrand}}, \bibinfo {author} {\bibfnamefont {A.}~\bibnamefont
  {Brillante}},\ and\ \bibinfo {author} {\bibfnamefont {D.}~\bibnamefont
  {M\"{o}bius}},\ }\href {https://doi.org/10.1063/1.443834} {\bibfield
  {journal} {\bibinfo  {journal} {The Journal of Chemical Physics}\ }\textbf
  {\bibinfo {volume} {77}},\ \bibinfo {pages} {6289} (\bibinfo {year}
  {1982})}\BibitemShut {NoStop}%
\bibitem [{\citenamefont {Lidzey}\ \emph {et~al.}(1998)\citenamefont {Lidzey},
  \citenamefont {Bradley}, \citenamefont {Skolnick}, \citenamefont {Virgili},
  \citenamefont {Walker},\ and\ \citenamefont {Whittaker}}]{Lidzey1998}%
  \BibitemOpen
  \bibfield  {author} {\bibinfo {author} {\bibfnamefont {D.~G.}\ \bibnamefont
  {Lidzey}}, \bibinfo {author} {\bibfnamefont {D.~D.~C.}\ \bibnamefont
  {Bradley}}, \bibinfo {author} {\bibfnamefont {M.~S.}\ \bibnamefont
  {Skolnick}}, \bibinfo {author} {\bibfnamefont {T.}~\bibnamefont {Virgili}},
  \bibinfo {author} {\bibfnamefont {S.}~\bibnamefont {Walker}},\ and\ \bibinfo
  {author} {\bibfnamefont {D.~M.}\ \bibnamefont {Whittaker}},\ }\href
  {https://doi.org/10.1038/25692} {\bibfield  {journal} {\bibinfo  {journal}
  {Nature}\ }\textbf {\bibinfo {volume} {395}},\ \bibinfo {pages} {53}
  (\bibinfo {year} {1998})}\BibitemShut {NoStop}%
\bibitem [{\citenamefont {Bellessa}\ \emph {et~al.}(2004)\citenamefont
  {Bellessa}, \citenamefont {Bonnand}, \citenamefont {Plenet},\ and\
  \citenamefont {Mugnier}}]{Bellessa2004}%
  \BibitemOpen
  \bibfield  {author} {\bibinfo {author} {\bibfnamefont {J.}~\bibnamefont
  {Bellessa}}, \bibinfo {author} {\bibfnamefont {C.}~\bibnamefont {Bonnand}},
  \bibinfo {author} {\bibfnamefont {J.~C.}\ \bibnamefont {Plenet}},\ and\
  \bibinfo {author} {\bibfnamefont {J.}~\bibnamefont {Mugnier}},\ }\href
  {https://doi.org/10.1103/PhysRevLett.93.036404} {\bibfield  {journal}
  {\bibinfo  {journal} {Phys. Rev. Lett.}\ }\textbf {\bibinfo {volume} {93}},\
  \bibinfo {pages} {036404} (\bibinfo {year} {2004})}\BibitemShut {NoStop}%
\bibitem [{\citenamefont {Dintinger}\ \emph {et~al.}(2005)\citenamefont
  {Dintinger}, \citenamefont {Klein}, \citenamefont {Bustos}, \citenamefont
  {Barnes},\ and\ \citenamefont {Ebbesen}}]{dintinger2005}%
  \BibitemOpen
  \bibfield  {author} {\bibinfo {author} {\bibfnamefont {J.}~\bibnamefont
  {Dintinger}}, \bibinfo {author} {\bibfnamefont {S.}~\bibnamefont {Klein}},
  \bibinfo {author} {\bibfnamefont {F.}~\bibnamefont {Bustos}}, \bibinfo
  {author} {\bibfnamefont {W.~L.}\ \bibnamefont {Barnes}},\ and\ \bibinfo
  {author} {\bibfnamefont {T.~W.}\ \bibnamefont {Ebbesen}},\ }\href
  {https://doi.org/10.1103/PhysRevB.71.035424} {\bibfield  {journal} {\bibinfo
  {journal} {Phys. Rev. B}\ }\textbf {\bibinfo {volume} {71}},\ \bibinfo
  {pages} {035424} (\bibinfo {year} {2005})}\BibitemShut {NoStop}%
\bibitem [{\citenamefont {Shalabney}\ \emph {et~al.}(2015)\citenamefont
  {Shalabney}, \citenamefont {George}, \citenamefont {Hutchison}, \citenamefont
  {Pupillo}, \citenamefont {Genet},\ and\ \citenamefont
  {Ebbesen}}]{Shalabney2015}%
  \BibitemOpen
  \bibfield  {author} {\bibinfo {author} {\bibfnamefont {A.}~\bibnamefont
  {Shalabney}}, \bibinfo {author} {\bibfnamefont {J.}~\bibnamefont {George}},
  \bibinfo {author} {\bibfnamefont {J.~A.}\ \bibnamefont {Hutchison}}, \bibinfo
  {author} {\bibfnamefont {G.}~\bibnamefont {Pupillo}}, \bibinfo {author}
  {\bibfnamefont {C.}~\bibnamefont {Genet}},\ and\ \bibinfo {author}
  {\bibfnamefont {T.~W.}\ \bibnamefont {Ebbesen}},\ }\href
  {https://doi.org/10.1038/ncomms6981} {\bibfield  {journal} {\bibinfo
  {journal} {Nature Communications}\ }\textbf {\bibinfo {volume} {6}},\
  \bibinfo {pages} {5981} (\bibinfo {year} {2015})}\BibitemShut {NoStop}%
\bibitem [{\citenamefont {Chikkaraddy}\ \emph {et~al.}(2016)\citenamefont
  {Chikkaraddy}, \citenamefont {de~Nijs}, \citenamefont {Benz}, \citenamefont
  {Barrow}, \citenamefont {Scherman}, \citenamefont {Rosta}, \citenamefont
  {Demetriadou}, \citenamefont {Fox}, \citenamefont {Hess},\ and\ \citenamefont
  {Baumberg}}]{Chikkaraddy2016}%
  \BibitemOpen
  \bibfield  {author} {\bibinfo {author} {\bibfnamefont {R.}~\bibnamefont
  {Chikkaraddy}}, \bibinfo {author} {\bibfnamefont {B.}~\bibnamefont
  {de~Nijs}}, \bibinfo {author} {\bibfnamefont {F.}~\bibnamefont {Benz}},
  \bibinfo {author} {\bibfnamefont {S.~J.}\ \bibnamefont {Barrow}}, \bibinfo
  {author} {\bibfnamefont {O.~A.}\ \bibnamefont {Scherman}}, \bibinfo {author}
  {\bibfnamefont {E.}~\bibnamefont {Rosta}}, \bibinfo {author} {\bibfnamefont
  {A.}~\bibnamefont {Demetriadou}}, \bibinfo {author} {\bibfnamefont
  {P.}~\bibnamefont {Fox}}, \bibinfo {author} {\bibfnamefont {O.}~\bibnamefont
  {Hess}},\ and\ \bibinfo {author} {\bibfnamefont {J.~J.}\ \bibnamefont
  {Baumberg}},\ }\href {https://doi.org/10.1038/nature17974} {\bibfield
  {journal} {\bibinfo  {journal} {Nature}\ }\textbf {\bibinfo {volume} {535}},\
  \bibinfo {pages} {127} (\bibinfo {year} {2016})}\BibitemShut {NoStop}%
\bibitem [{\citenamefont {Wang}\ \emph {et~al.}(2017)\citenamefont {Wang},
  \citenamefont {Kelkar}, \citenamefont {Martin-Cano}, \citenamefont {Utikal},
  \citenamefont {G\"otzinger},\ and\ \citenamefont {Sandoghdar}}]{Wang2017}%
  \BibitemOpen
  \bibfield  {author} {\bibinfo {author} {\bibfnamefont {D.}~\bibnamefont
  {Wang}}, \bibinfo {author} {\bibfnamefont {H.}~\bibnamefont {Kelkar}},
  \bibinfo {author} {\bibfnamefont {D.}~\bibnamefont {Martin-Cano}}, \bibinfo
  {author} {\bibfnamefont {T.}~\bibnamefont {Utikal}}, \bibinfo {author}
  {\bibfnamefont {S.}~\bibnamefont {G\"otzinger}},\ and\ \bibinfo {author}
  {\bibfnamefont {V.}~\bibnamefont {Sandoghdar}},\ }\href
  {https://doi.org/10.1103/PhysRevX.7.021014} {\bibfield  {journal} {\bibinfo
  {journal} {Phys. Rev. X}\ }\textbf {\bibinfo {volume} {7}},\ \bibinfo {pages}
  {021014} (\bibinfo {year} {2017})}\BibitemShut {NoStop}%
\bibitem [{\citenamefont {Zhang}\ \emph {et~al.}(2017)\citenamefont {Zhang},
  \citenamefont {Meng}, \citenamefont {Zhang}, \citenamefont {Luo},
  \citenamefont {Yu}, \citenamefont {Yang}, \citenamefont {Zhang},
  \citenamefont {Esteban}, \citenamefont {Aizpurua}, \citenamefont {Luo},
  \citenamefont {Yang}, \citenamefont {Dong},\ and\ \citenamefont
  {Hou}}]{Zhang2017}%
  \BibitemOpen
  \bibfield  {author} {\bibinfo {author} {\bibfnamefont {Y.}~\bibnamefont
  {Zhang}}, \bibinfo {author} {\bibfnamefont {Q.-S.}\ \bibnamefont {Meng}},
  \bibinfo {author} {\bibfnamefont {L.}~\bibnamefont {Zhang}}, \bibinfo
  {author} {\bibfnamefont {Y.}~\bibnamefont {Luo}}, \bibinfo {author}
  {\bibfnamefont {Y.-J.}\ \bibnamefont {Yu}}, \bibinfo {author} {\bibfnamefont
  {B.}~\bibnamefont {Yang}}, \bibinfo {author} {\bibfnamefont {Y.}~\bibnamefont
  {Zhang}}, \bibinfo {author} {\bibfnamefont {R.}~\bibnamefont {Esteban}},
  \bibinfo {author} {\bibfnamefont {J.}~\bibnamefont {Aizpurua}}, \bibinfo
  {author} {\bibfnamefont {Y.}~\bibnamefont {Luo}}, \bibinfo {author}
  {\bibfnamefont {J.-L.}\ \bibnamefont {Yang}}, \bibinfo {author}
  {\bibfnamefont {Z.-C.}\ \bibnamefont {Dong}},\ and\ \bibinfo {author}
  {\bibfnamefont {J.~G.}\ \bibnamefont {Hou}},\ }\href
  {http://dx.doi.org/10.1038/ncomms15225} {\bibfield  {journal} {\bibinfo
  {journal} {Nature Communications}\ }\textbf {\bibinfo {volume} {8}},\
  \bibinfo {pages} {15225} (\bibinfo {year} {2017})}\BibitemShut {NoStop}%
\bibitem [{\citenamefont {Lather}\ \emph {et~al.}(2019)\citenamefont {Lather},
  \citenamefont {Bhatt}, \citenamefont {Thomas}, \citenamefont {Ebbesen},\ and\
  \citenamefont {George}}]{Lather2018}%
  \BibitemOpen
  \bibfield  {author} {\bibinfo {author} {\bibfnamefont {J.}~\bibnamefont
  {Lather}}, \bibinfo {author} {\bibfnamefont {P.}~\bibnamefont {Bhatt}},
  \bibinfo {author} {\bibfnamefont {A.}~\bibnamefont {Thomas}}, \bibinfo
  {author} {\bibfnamefont {T.~W.}\ \bibnamefont {Ebbesen}},\ and\ \bibinfo
  {author} {\bibfnamefont {J.}~\bibnamefont {George}},\ }\href
  {https://doi.org/10.1002/anie.201905407} {\bibfield  {journal} {\bibinfo
  {journal} {Angewandte Chemie International Edition}\ }\textbf {\bibinfo
  {volume} {58}},\ \bibinfo {pages} {10635} (\bibinfo {year}
  {2019})}\BibitemShut {NoStop}%
\bibitem [{\citenamefont {Thomas}\ \emph
  {et~al.}(2019{\natexlab{a}})\citenamefont {Thomas}, \citenamefont {Devaux},
  \citenamefont {Nagarajan}, \citenamefont {Chervy}, \citenamefont {Seidel},
  \citenamefont {Hagenm{\ifmmode\ddot{u}\else\"{u}\fi}ller}, \citenamefont
  {Sch{\ifmmode\ddot{u}\else\"{u}\fi}tz}, \citenamefont {Schachenmayer},
  \citenamefont {Genet}, \citenamefont {Pupillo},\ and\ \citenamefont
  {Ebbesen}}]{Thomas_Explor_2019}%
  \BibitemOpen
  \bibfield  {author} {\bibinfo {author} {\bibfnamefont {A.}~\bibnamefont
  {Thomas}}, \bibinfo {author} {\bibfnamefont {E.}~\bibnamefont {Devaux}},
  \bibinfo {author} {\bibfnamefont {K.}~\bibnamefont {Nagarajan}}, \bibinfo
  {author} {\bibfnamefont {T.}~\bibnamefont {Chervy}}, \bibinfo {author}
  {\bibfnamefont {M.}~\bibnamefont {Seidel}}, \bibinfo {author} {\bibfnamefont
  {D.}~\bibnamefont {Hagenm{\ifmmode\ddot{u}\else\"{u}\fi}ller}}, \bibinfo
  {author} {\bibfnamefont {S.}~\bibnamefont
  {Sch{\ifmmode\ddot{u}\else\"{u}\fi}tz}}, \bibinfo {author} {\bibfnamefont
  {J.}~\bibnamefont {Schachenmayer}}, \bibinfo {author} {\bibfnamefont
  {C.}~\bibnamefont {Genet}}, \bibinfo {author} {\bibfnamefont
  {G.}~\bibnamefont {Pupillo}},\ and\ \bibinfo {author} {\bibfnamefont {T.~W.}\
  \bibnamefont {Ebbesen}},\ }\href {https://arxiv.org/abs/1911.01459}
  {\bibfield  {journal} {\bibinfo  {journal} {arXiv}\ } (\bibinfo {year}
  {2019}{\natexlab{a}})},\ \Eprint {https://arxiv.org/abs/1911.01459}
  {1911.01459} \BibitemShut {NoStop}%
\bibitem [{\citenamefont {Park}\ \emph {et~al.}(2006)\citenamefont {Park},
  \citenamefont {Cook},\ and\ \citenamefont {Wang}}]{Park2006}%
  \BibitemOpen
  \bibfield  {author} {\bibinfo {author} {\bibfnamefont {Y.-S.}\ \bibnamefont
  {Park}}, \bibinfo {author} {\bibfnamefont {A.~K.}\ \bibnamefont {Cook}},\
  and\ \bibinfo {author} {\bibfnamefont {H.}~\bibnamefont {Wang}},\ }\href
  {https://doi.org/10.1021/nl061342r} {\bibfield  {journal} {\bibinfo
  {journal} {Nano Letters}\ }\textbf {\bibinfo {volume} {6}},\ \bibinfo {pages}
  {2075} (\bibinfo {year} {2006})}\BibitemShut {NoStop}%
\bibitem [{\citenamefont {Bradac}\ \emph {et~al.}(2017)\citenamefont {Bradac},
  \citenamefont {Johnsson}, \citenamefont {van Breugel}, \citenamefont
  {Baragiola}, \citenamefont {Martin}, \citenamefont {Juan}, \citenamefont
  {Brennen},\ and\ \citenamefont {Volz}}]{Bradac2017}%
  \BibitemOpen
  \bibfield  {author} {\bibinfo {author} {\bibfnamefont {C.}~\bibnamefont
  {Bradac}}, \bibinfo {author} {\bibfnamefont {M.~T.}\ \bibnamefont
  {Johnsson}}, \bibinfo {author} {\bibfnamefont {M.}~\bibnamefont {van
  Breugel}}, \bibinfo {author} {\bibfnamefont {B.~Q.}\ \bibnamefont
  {Baragiola}}, \bibinfo {author} {\bibfnamefont {R.}~\bibnamefont {Martin}},
  \bibinfo {author} {\bibfnamefont {M.~L.}\ \bibnamefont {Juan}}, \bibinfo
  {author} {\bibfnamefont {G.~K.}\ \bibnamefont {Brennen}},\ and\ \bibinfo
  {author} {\bibfnamefont {T.}~\bibnamefont {Volz}},\ }\href
  {https://doi.org/10.1038/s41467-017-01397-4} {\bibfield  {journal} {\bibinfo
  {journal} {Nature Communications}\ }\textbf {\bibinfo {volume} {8}},\
  \bibinfo {pages} {1205} (\bibinfo {year} {2017})}\BibitemShut {NoStop}%
\bibitem [{\citenamefont {Astner}\ \emph {et~al.}(2017)\citenamefont {Astner},
  \citenamefont {Nevlacsil}, \citenamefont {Peterschofsky}, \citenamefont
  {Angerer}, \citenamefont {Rotter}, \citenamefont {Putz}, \citenamefont
  {Schmiedmayer},\ and\ \citenamefont {Majer}}]{Astner2017}%
  \BibitemOpen
  \bibfield  {author} {\bibinfo {author} {\bibfnamefont {T.}~\bibnamefont
  {Astner}}, \bibinfo {author} {\bibfnamefont {S.}~\bibnamefont {Nevlacsil}},
  \bibinfo {author} {\bibfnamefont {N.}~\bibnamefont {Peterschofsky}}, \bibinfo
  {author} {\bibfnamefont {A.}~\bibnamefont {Angerer}}, \bibinfo {author}
  {\bibfnamefont {S.}~\bibnamefont {Rotter}}, \bibinfo {author} {\bibfnamefont
  {S.}~\bibnamefont {Putz}}, \bibinfo {author} {\bibfnamefont {J.}~\bibnamefont
  {Schmiedmayer}},\ and\ \bibinfo {author} {\bibfnamefont {J.}~\bibnamefont
  {Majer}},\ }\href {https://doi.org/10.1103/PhysRevLett.118.140502} {\bibfield
   {journal} {\bibinfo  {journal} {Phys. Rev. Lett.}\ }\textbf {\bibinfo
  {volume} {118}},\ \bibinfo {pages} {140502} (\bibinfo {year}
  {2017})}\BibitemShut {NoStop}%
\bibitem [{\citenamefont {Angerer}\ \emph {et~al.}(2018)\citenamefont
  {Angerer}, \citenamefont {Streltsov}, \citenamefont {Astner}, \citenamefont
  {Putz}, \citenamefont {Sumiya}, \citenamefont {Onoda}, \citenamefont {Isoya},
  \citenamefont {Munro}, \citenamefont {Nemoto}, \citenamefont {Schmiedmayer},\
  and\ \citenamefont {Majer}}]{Angerer2018}%
  \BibitemOpen
  \bibfield  {author} {\bibinfo {author} {\bibfnamefont {A.}~\bibnamefont
  {Angerer}}, \bibinfo {author} {\bibfnamefont {K.}~\bibnamefont {Streltsov}},
  \bibinfo {author} {\bibfnamefont {T.}~\bibnamefont {Astner}}, \bibinfo
  {author} {\bibfnamefont {S.}~\bibnamefont {Putz}}, \bibinfo {author}
  {\bibfnamefont {H.}~\bibnamefont {Sumiya}}, \bibinfo {author} {\bibfnamefont
  {S.}~\bibnamefont {Onoda}}, \bibinfo {author} {\bibfnamefont
  {J.}~\bibnamefont {Isoya}}, \bibinfo {author} {\bibfnamefont {W.~J.}\
  \bibnamefont {Munro}}, \bibinfo {author} {\bibfnamefont {K.}~\bibnamefont
  {Nemoto}}, \bibinfo {author} {\bibfnamefont {J.}~\bibnamefont
  {Schmiedmayer}},\ and\ \bibinfo {author} {\bibfnamefont {J.}~\bibnamefont
  {Majer}},\ }\href {https://doi.org/10.1038/s41567-018-0269-7} {\bibfield
  {journal} {\bibinfo  {journal} {Nature Physics}\ }\textbf {\bibinfo {volume}
  {14}},\ \bibinfo {pages} {1168} (\bibinfo {year} {2018})}\BibitemShut
  {NoStop}%
\bibitem [{\citenamefont {Evans}\ \emph {et~al.}(2018)\citenamefont {Evans},
  \citenamefont {Bhaskar}, \citenamefont {Sukachev}, \citenamefont {Nguyen},
  \citenamefont {Sipahigil}, \citenamefont {Burek}, \citenamefont {Machielse},
  \citenamefont {Zhang}, \citenamefont {Zibrov}, \citenamefont {Bielejec},
  \citenamefont {Park}, \citenamefont {Lon\v{c}ar},\ and\ \citenamefont
  {Lukin}}]{Evans2018}%
  \BibitemOpen
  \bibfield  {author} {\bibinfo {author} {\bibfnamefont {R.~E.}\ \bibnamefont
  {Evans}}, \bibinfo {author} {\bibfnamefont {M.~K.}\ \bibnamefont {Bhaskar}},
  \bibinfo {author} {\bibfnamefont {D.~D.}\ \bibnamefont {Sukachev}}, \bibinfo
  {author} {\bibfnamefont {C.~T.}\ \bibnamefont {Nguyen}}, \bibinfo {author}
  {\bibfnamefont {A.}~\bibnamefont {Sipahigil}}, \bibinfo {author}
  {\bibfnamefont {M.~J.}\ \bibnamefont {Burek}}, \bibinfo {author}
  {\bibfnamefont {B.}~\bibnamefont {Machielse}}, \bibinfo {author}
  {\bibfnamefont {G.~H.}\ \bibnamefont {Zhang}}, \bibinfo {author}
  {\bibfnamefont {A.~S.}\ \bibnamefont {Zibrov}}, \bibinfo {author}
  {\bibfnamefont {E.}~\bibnamefont {Bielejec}}, \bibinfo {author}
  {\bibfnamefont {H.}~\bibnamefont {Park}}, \bibinfo {author} {\bibfnamefont
  {M.}~\bibnamefont {Lon\v{c}ar}},\ and\ \bibinfo {author} {\bibfnamefont
  {M.~D.}\ \bibnamefont {Lukin}},\ }\href
  {https://doi.org/10.1126/science.aau4691} {\bibfield  {journal} {\bibinfo
  {journal} {Science}\ }\textbf {\bibinfo {volume} {362}},\ \bibinfo {pages}
  {662} (\bibinfo {year} {2018})}\BibitemShut {NoStop}%
\bibitem [{\citenamefont {Liu}\ \emph {et~al.}(2014)\citenamefont {Liu},
  \citenamefont {Luan}, \citenamefont {Li}, \citenamefont {Gong}, \citenamefont
  {Wong},\ and\ \citenamefont {Xiao}}]{Liu_Cohere_2014}%
  \BibitemOpen
  \bibfield  {author} {\bibinfo {author} {\bibfnamefont {Y.-C.}\ \bibnamefont
  {Liu}}, \bibinfo {author} {\bibfnamefont {X.}~\bibnamefont {Luan}}, \bibinfo
  {author} {\bibfnamefont {H.-K.}\ \bibnamefont {Li}}, \bibinfo {author}
  {\bibfnamefont {Q.}~\bibnamefont {Gong}}, \bibinfo {author} {\bibfnamefont
  {C.~W.}\ \bibnamefont {Wong}},\ and\ \bibinfo {author} {\bibfnamefont
  {Y.-F.}\ \bibnamefont {Xiao}},\ }\href
  {https://doi.org/10.1103/PhysRevLett.112.213602} {\bibfield  {journal}
  {\bibinfo  {journal} {Phys. Rev. Lett.}\ }\textbf {\bibinfo {volume} {112}},\
  \bibinfo {pages} {213602} (\bibinfo {year} {2014})}\BibitemShut {NoStop}%
\bibitem [{\citenamefont
  {Ram{\ifmmode\acute{\imath}\else\'{\i}\fi}rez-Mu{\ifmmode\tilde{n}\else\~{n}\fi}oz}\
  \emph {et~al.}(2018)\citenamefont
  {Ram{\ifmmode\acute{\imath}\else\'{\i}\fi}rez-Mu{\ifmmode\tilde{n}\else\~{n}\fi}oz},
  \citenamefont {Restrepo~Cuartas},\ and\ \citenamefont
  {Vinck-Posada}}]{Ramirez-Munoz_Indire_2018}%
  \BibitemOpen
  \bibfield  {author} {\bibinfo {author} {\bibfnamefont {J.~E.}\ \bibnamefont
  {Ram{\ifmmode\acute{\imath}\else\'{\i}\fi}rez-Mu{\ifmmode\tilde{n}\else\~{n}\fi}oz}},
  \bibinfo {author} {\bibfnamefont {J.~P.}\ \bibnamefont {Restrepo~Cuartas}},\
  and\ \bibinfo {author} {\bibfnamefont {H.}~\bibnamefont {Vinck-Posada}},\
  }\href {https://doi.org/10.1016/j.physleta.2018.08.001} {\bibfield  {journal}
  {\bibinfo  {journal} {Phys. Lett. A}\ }\textbf {\bibinfo {volume} {382}},\
  \bibinfo {pages} {3109} (\bibinfo {year} {2018})}\BibitemShut {NoStop}%
\bibitem [{\citenamefont {Reiter}\ and\ \citenamefont
  {S{\o}rensen}(2012)}]{Reiter_Effect_2012}%
  \BibitemOpen
  \bibfield  {author} {\bibinfo {author} {\bibfnamefont {F.}~\bibnamefont
  {Reiter}}\ and\ \bibinfo {author} {\bibfnamefont {A.~S.}\ \bibnamefont
  {S{\o}rensen}},\ }\href {https://doi.org/10.1103/PhysRevA.85.032111}
  {\bibfield  {journal} {\bibinfo  {journal} {Phys. Rev. A}\ }\textbf {\bibinfo
  {volume} {85}},\ \bibinfo {pages} {032111} (\bibinfo {year}
  {2012})}\BibitemShut {NoStop}%
\bibitem [{\citenamefont {Poyatos}\ \emph {et~al.}(1996)\citenamefont
  {Poyatos}, \citenamefont {Cirac},\ and\ \citenamefont
  {Zoller}}]{Poyatos_Quantu_1996}%
  \BibitemOpen
  \bibfield  {author} {\bibinfo {author} {\bibfnamefont {J.~F.}\ \bibnamefont
  {Poyatos}}, \bibinfo {author} {\bibfnamefont {J.~I.}\ \bibnamefont {Cirac}},\
  and\ \bibinfo {author} {\bibfnamefont {P.}~\bibnamefont {Zoller}},\ }\href
  {https://doi.org/10.1103/PhysRevLett.77.4728} {\bibfield  {journal} {\bibinfo
   {journal} {Phys. Rev. Lett.}\ }\textbf {\bibinfo {volume} {77}},\ \bibinfo
  {pages} {4728} (\bibinfo {year} {1996})}\BibitemShut {NoStop}%
\bibitem [{\citenamefont {Kraus}\ \emph {et~al.}(2008)\citenamefont {Kraus},
  \citenamefont {B\"uchler}, \citenamefont {Diehl}, \citenamefont {Kantian},
  \citenamefont {Micheli},\ and\ \citenamefont {Zoller}}]{Kraus_Prepar_2008}%
  \BibitemOpen
  \bibfield  {author} {\bibinfo {author} {\bibfnamefont {B.}~\bibnamefont
  {Kraus}}, \bibinfo {author} {\bibfnamefont {H.~P.}\ \bibnamefont
  {B\"uchler}}, \bibinfo {author} {\bibfnamefont {S.}~\bibnamefont {Diehl}},
  \bibinfo {author} {\bibfnamefont {A.}~\bibnamefont {Kantian}}, \bibinfo
  {author} {\bibfnamefont {A.}~\bibnamefont {Micheli}},\ and\ \bibinfo {author}
  {\bibfnamefont {P.}~\bibnamefont {Zoller}},\ }\href
  {https://doi.org/10.1103/PhysRevA.78.042307} {\bibfield  {journal} {\bibinfo
  {journal} {Phys. Rev. A}\ }\textbf {\bibinfo {volume} {78}},\ \bibinfo
  {pages} {042307} (\bibinfo {year} {2008})}\BibitemShut {NoStop}%
\bibitem [{\citenamefont {Muschik}\ \emph {et~al.}(2011)\citenamefont
  {Muschik}, \citenamefont {Polzik},\ and\ \citenamefont
  {Cirac}}]{Muschik_Dissip_2011}%
  \BibitemOpen
  \bibfield  {author} {\bibinfo {author} {\bibfnamefont {C.~A.}\ \bibnamefont
  {Muschik}}, \bibinfo {author} {\bibfnamefont {E.~S.}\ \bibnamefont
  {Polzik}},\ and\ \bibinfo {author} {\bibfnamefont {J.~I.}\ \bibnamefont
  {Cirac}},\ }\href {https://doi.org/10.1103/PhysRevA.83.052312} {\bibfield
  {journal} {\bibinfo  {journal} {Phys. Rev. A}\ }\textbf {\bibinfo {volume}
  {83}},\ \bibinfo {pages} {052312} (\bibinfo {year} {2011})}\BibitemShut
  {NoStop}%
\bibitem [{\citenamefont {Wang}\ and\ \citenamefont
  {Clerk}(2013)}]{Wang_Reserv_2013}%
  \BibitemOpen
  \bibfield  {author} {\bibinfo {author} {\bibfnamefont {Y.-D.}\ \bibnamefont
  {Wang}}\ and\ \bibinfo {author} {\bibfnamefont {A.~A.}\ \bibnamefont
  {Clerk}},\ }\href {https://doi.org/10.1103/PhysRevLett.110.253601} {\bibfield
   {journal} {\bibinfo  {journal} {Phys. Rev. Lett.}\ }\textbf {\bibinfo
  {volume} {110}},\ \bibinfo {pages} {253601} (\bibinfo {year}
  {2013})}\BibitemShut {NoStop}%
\bibitem [{\citenamefont {Zanardi}\ and\ \citenamefont
  {Campos~Venuti}(2014)}]{Zanardi_Cohere_2014}%
  \BibitemOpen
  \bibfield  {author} {\bibinfo {author} {\bibfnamefont {P.}~\bibnamefont
  {Zanardi}}\ and\ \bibinfo {author} {\bibfnamefont {L.}~\bibnamefont
  {Campos~Venuti}},\ }\href {https://doi.org/10.1103/PhysRevLett.113.240406}
  {\bibfield  {journal} {\bibinfo  {journal} {Phys. Rev. Lett.}\ }\textbf
  {\bibinfo {volume} {113}},\ \bibinfo {pages} {240406} (\bibinfo {year}
  {2014})}\BibitemShut {NoStop}%
\bibitem [{\citenamefont {Arenz}\ and\ \citenamefont
  {Metelmann}(2019)}]{Arenz_Emergi_2019}%
  \BibitemOpen
  \bibfield  {author} {\bibinfo {author} {\bibfnamefont {C.}~\bibnamefont
  {Arenz}}\ and\ \bibinfo {author} {\bibfnamefont {A.}~\bibnamefont
  {Metelmann}},\ }\href {https://arxiv.org/abs/1906.03481} {\bibfield
  {journal} {\bibinfo  {journal} {arXiv}\ } (\bibinfo {year} {2019})},\ \Eprint
  {https://arxiv.org/abs/1906.03481} {1906.03481} \BibitemShut {NoStop}%
\bibitem [{\citenamefont {Canaguier-Durand}\ \emph {et~al.}(2015)\citenamefont
  {Canaguier-Durand}, \citenamefont {Genet}, \citenamefont {Lambrecht},
  \citenamefont {Ebbesen},\ and\ \citenamefont
  {Reynaud}}]{Canaguier-Durand_Non-Ma_2015}%
  \BibitemOpen
  \bibfield  {author} {\bibinfo {author} {\bibfnamefont {A.}~\bibnamefont
  {Canaguier-Durand}}, \bibinfo {author} {\bibfnamefont {C.}~\bibnamefont
  {Genet}}, \bibinfo {author} {\bibfnamefont {A.}~\bibnamefont {Lambrecht}},
  \bibinfo {author} {\bibfnamefont {T.~W.}\ \bibnamefont {Ebbesen}},\ and\
  \bibinfo {author} {\bibfnamefont {S.}~\bibnamefont {Reynaud}},\ }\href
  {https://doi.org/10.1140/epjd/e2014-50539-x} {\bibfield  {journal} {\bibinfo
  {journal} {Eur. Phys. J. D}\ }\textbf {\bibinfo {volume} {69}},\ \bibinfo
  {pages} {24} (\bibinfo {year} {2015})}\BibitemShut {NoStop}%
\bibitem [{\citenamefont {Neuman}\ and\ \citenamefont
  {Aizpurua}(2018)}]{Neuman_Origin_2018}%
  \BibitemOpen
  \bibfield  {author} {\bibinfo {author} {\bibfnamefont {T.}~\bibnamefont
  {Neuman}}\ and\ \bibinfo {author} {\bibfnamefont {J.}~\bibnamefont
  {Aizpurua}},\ }\href {https://doi.org/10.1364/OPTICA.5.001247} {\bibfield
  {journal} {\bibinfo  {journal} {Optica}\ }\textbf {\bibinfo {volume} {5}},\
  \bibinfo {pages} {1247} (\bibinfo {year} {2018})}\BibitemShut {NoStop}%
\bibitem [{\citenamefont {Jaynes}\ and\ \citenamefont
  {Cummings}(1963)}]{Jaynes_Compar_1963}%
  \BibitemOpen
  \bibfield  {author} {\bibinfo {author} {\bibfnamefont {E.~T.}\ \bibnamefont
  {Jaynes}}\ and\ \bibinfo {author} {\bibfnamefont {F.~W.}\ \bibnamefont
  {Cummings}},\ }\href {https://doi.org/10.1109/PROC.1963.1664} {\bibfield
  {journal} {\bibinfo  {journal} {Proceedings of the IEEE}\ }\textbf {\bibinfo
  {volume} {51}},\ \bibinfo {pages} {89} (\bibinfo {year} {1963})}\BibitemShut
  {NoStop}%
\bibitem [{\citenamefont {Tavis}\ and\ \citenamefont
  {Cummings}(1968)}]{Tavis_Exact_1968}%
  \BibitemOpen
  \bibfield  {author} {\bibinfo {author} {\bibfnamefont {M.}~\bibnamefont
  {Tavis}}\ and\ \bibinfo {author} {\bibfnamefont {F.~W.}\ \bibnamefont
  {Cummings}},\ }\href {https://doi.org/10.1103/PhysRev.170.379} {\bibfield
  {journal} {\bibinfo  {journal} {Phys. Rev.}\ }\textbf {\bibinfo {volume}
  {170}},\ \bibinfo {pages} {379} (\bibinfo {year} {1968})}\BibitemShut
  {NoStop}%
\bibitem [{\citenamefont {Lehmberg}(1970)}]{Lehmberg1970_Radia}%
  \BibitemOpen
  \bibfield  {author} {\bibinfo {author} {\bibfnamefont {R.~H.}\ \bibnamefont
  {Lehmberg}},\ }\href {https://doi.org/10.1103/PhysRevA.2.883} {\bibfield
  {journal} {\bibinfo  {journal} {Phys. Rev. A}\ }\textbf {\bibinfo {volume}
  {2}},\ \bibinfo {pages} {883} (\bibinfo {year} {1970})}\BibitemShut {NoStop}%
\bibitem [{\citenamefont {Zwanzig}(1960)}]{Zwanzig1960}%
  \BibitemOpen
  \bibfield  {author} {\bibinfo {author} {\bibfnamefont {R.}~\bibnamefont
  {Zwanzig}},\ }\href {https://doi.org/10.1063/1.1731409} {\bibfield  {journal}
  {\bibinfo  {journal} {The Journal of Chemical Physics}\ }\textbf {\bibinfo
  {volume} {33}},\ \bibinfo {pages} {1338} (\bibinfo {year}
  {1960})}\BibitemShut {NoStop}%
\bibitem [{\citenamefont {Bonifacio}\ \emph {et~al.}(1971)\citenamefont
  {Bonifacio}, \citenamefont {Schwendimann},\ and\ \citenamefont
  {Haake}}]{Bonifacio1971}%
  \BibitemOpen
  \bibfield  {author} {\bibinfo {author} {\bibfnamefont {R.}~\bibnamefont
  {Bonifacio}}, \bibinfo {author} {\bibfnamefont {P.}~\bibnamefont
  {Schwendimann}},\ and\ \bibinfo {author} {\bibfnamefont {F.}~\bibnamefont
  {Haake}},\ }\href {https://doi.org/10.1103/PhysRevA.4.302} {\bibfield
  {journal} {\bibinfo  {journal} {Phys. Rev. A}\ }\textbf {\bibinfo {volume}
  {4}},\ \bibinfo {pages} {302} (\bibinfo {year} {1971})}\BibitemShut {NoStop}%
\bibitem [{\citenamefont {Sch\"utz}\ \emph {et~al.}(2013)\citenamefont
  {Sch\"utz}, \citenamefont {Habibian},\ and\ \citenamefont
  {Morigi}}]{Schuetz2013}%
  \BibitemOpen
  \bibfield  {author} {\bibinfo {author} {\bibfnamefont {S.}~\bibnamefont
  {Sch\"utz}}, \bibinfo {author} {\bibfnamefont {H.}~\bibnamefont {Habibian}},\
  and\ \bibinfo {author} {\bibfnamefont {G.}~\bibnamefont {Morigi}},\ }\href
  {https://doi.org/10.1103/PhysRevA.88.033427} {\bibfield  {journal} {\bibinfo
  {journal} {Phys. Rev. A}\ }\textbf {\bibinfo {volume} {88}},\ \bibinfo
  {pages} {033427} (\bibinfo {year} {2013})}\BibitemShut {NoStop}%
\bibitem [{\citenamefont {Hagenm\"uller}\ \emph {et~al.}(2018)\citenamefont
  {Hagenm\"uller}, \citenamefont {Sch\"utz}, \citenamefont {Schachenmayer},
  \citenamefont {Genes},\ and\ \citenamefont {Pupillo}}]{PhysRevB.97.205303}%
  \BibitemOpen
  \bibfield  {author} {\bibinfo {author} {\bibfnamefont {D.}~\bibnamefont
  {Hagenm\"uller}}, \bibinfo {author} {\bibfnamefont {S.}~\bibnamefont
  {Sch\"utz}}, \bibinfo {author} {\bibfnamefont {J.}~\bibnamefont
  {Schachenmayer}}, \bibinfo {author} {\bibfnamefont {C.}~\bibnamefont
  {Genes}},\ and\ \bibinfo {author} {\bibfnamefont {G.}~\bibnamefont
  {Pupillo}},\ }\href {https://doi.org/10.1103/PhysRevB.97.205303} {\bibfield
  {journal} {\bibinfo  {journal} {Phys. Rev. B}\ }\textbf {\bibinfo {volume}
  {97}},\ \bibinfo {pages} {205303} (\bibinfo {year} {2018})}\BibitemShut
  {NoStop}%
\bibitem [{\citenamefont {Horn}\ and\ \citenamefont
  {Johnson}(2012)}]{Horn2012}%
  \BibitemOpen
  \bibfield  {author} {\bibinfo {author} {\bibfnamefont {R.~A.}\ \bibnamefont
  {Horn}}\ and\ \bibinfo {author} {\bibfnamefont {C.~R.}\ \bibnamefont
  {Johnson}},\ }\href@noop {} {\emph {\bibinfo {title} {Matrix Analysis}}},\
  \bibinfo {edition} {2nd}\ ed.\ (\bibinfo  {publisher} {Cambridge University
  Press},\ \bibinfo {address} {New York, NY, USA},\ \bibinfo {year}
  {2012})\BibitemShut {NoStop}%
\bibitem [{\citenamefont {Carmichael}\ \emph {et~al.}(1989)\citenamefont
  {Carmichael}, \citenamefont {Brecha}, \citenamefont {Raizen}, \citenamefont
  {Kimble},\ and\ \citenamefont {Rice}}]{Carmichael1989}%
  \BibitemOpen
  \bibfield  {author} {\bibinfo {author} {\bibfnamefont {H.~J.}\ \bibnamefont
  {Carmichael}}, \bibinfo {author} {\bibfnamefont {R.~J.}\ \bibnamefont
  {Brecha}}, \bibinfo {author} {\bibfnamefont {M.~G.}\ \bibnamefont {Raizen}},
  \bibinfo {author} {\bibfnamefont {H.~J.}\ \bibnamefont {Kimble}},\ and\
  \bibinfo {author} {\bibfnamefont {P.~R.}\ \bibnamefont {Rice}},\ }\href
  {https://doi.org/10.1103/PhysRevA.40.5516} {\bibfield  {journal} {\bibinfo
  {journal} {Phys. Rev. A}\ }\textbf {\bibinfo {volume} {40}},\ \bibinfo
  {pages} {5516} (\bibinfo {year} {1989})}\BibitemShut {NoStop}%
\bibitem [{\citenamefont {James}(1993)}]{James1993}%
  \BibitemOpen
  \bibfield  {author} {\bibinfo {author} {\bibfnamefont {D.~F.~V.}\
  \bibnamefont {James}},\ }\href {https://doi.org/10.1103/PhysRevA.47.1336}
  {\bibfield  {journal} {\bibinfo  {journal} {Phys. Rev. A}\ }\textbf {\bibinfo
  {volume} {47}},\ \bibinfo {pages} {1336} (\bibinfo {year}
  {1993})}\BibitemShut {NoStop}%
\bibitem [{\citenamefont {Shammah}\ \emph {et~al.}(2018)\citenamefont
  {Shammah}, \citenamefont {Ahmed}, \citenamefont {Lambert}, \citenamefont
  {De~Liberato},\ and\ \citenamefont {Nori}}]{Shammah2018}%
  \BibitemOpen
  \bibfield  {author} {\bibinfo {author} {\bibfnamefont {N.}~\bibnamefont
  {Shammah}}, \bibinfo {author} {\bibfnamefont {S.}~\bibnamefont {Ahmed}},
  \bibinfo {author} {\bibfnamefont {N.}~\bibnamefont {Lambert}}, \bibinfo
  {author} {\bibfnamefont {S.}~\bibnamefont {De~Liberato}},\ and\ \bibinfo
  {author} {\bibfnamefont {F.}~\bibnamefont {Nori}},\ }\href
  {https://doi.org/10.1103/PhysRevA.98.063815} {\bibfield  {journal} {\bibinfo
  {journal} {Phys. Rev. A}\ }\textbf {\bibinfo {volume} {98}},\ \bibinfo
  {pages} {063815} (\bibinfo {year} {2018})}\BibitemShut {NoStop}%
\bibitem [{\citenamefont {Berry}(2004)}]{Berry2004}%
  \BibitemOpen
  \bibfield  {author} {\bibinfo {author} {\bibfnamefont {M.}~\bibnamefont
  {Berry}},\ }\href {https://doi.org/10.1023/B:CJOP.0000044002.05657.04}
  {\bibfield  {journal} {\bibinfo  {journal} {Czechoslovak Journal of Physics}\
  }\textbf {\bibinfo {volume} {54}},\ \bibinfo {pages} {1039} (\bibinfo {year}
  {2004})}\BibitemShut {NoStop}%
\bibitem [{\citenamefont {Gao}\ \emph {et~al.}(2018)\citenamefont {Gao},
  \citenamefont {Li}, \citenamefont {Estrecho}, \citenamefont {Liew},
  \citenamefont {Comber-Todd}, \citenamefont {Nalitov}, \citenamefont {Steger},
  \citenamefont {West}, \citenamefont {Pfeiffer}, \citenamefont {Snoke},
  \citenamefont {Kavokin}, \citenamefont {Truscott},\ and\ \citenamefont
  {Ostrovskaya}}]{Gao2018}%
  \BibitemOpen
  \bibfield  {author} {\bibinfo {author} {\bibfnamefont {T.}~\bibnamefont
  {Gao}}, \bibinfo {author} {\bibfnamefont {G.}~\bibnamefont {Li}}, \bibinfo
  {author} {\bibfnamefont {E.}~\bibnamefont {Estrecho}}, \bibinfo {author}
  {\bibfnamefont {T.~C.~H.}\ \bibnamefont {Liew}}, \bibinfo {author}
  {\bibfnamefont {D.}~\bibnamefont {Comber-Todd}}, \bibinfo {author}
  {\bibfnamefont {A.}~\bibnamefont {Nalitov}}, \bibinfo {author} {\bibfnamefont
  {M.}~\bibnamefont {Steger}}, \bibinfo {author} {\bibfnamefont
  {K.}~\bibnamefont {West}}, \bibinfo {author} {\bibfnamefont {L.}~\bibnamefont
  {Pfeiffer}}, \bibinfo {author} {\bibfnamefont {D.~W.}\ \bibnamefont {Snoke}},
  \bibinfo {author} {\bibfnamefont {A.~V.}\ \bibnamefont {Kavokin}}, \bibinfo
  {author} {\bibfnamefont {A.~G.}\ \bibnamefont {Truscott}},\ and\ \bibinfo
  {author} {\bibfnamefont {E.~A.}\ \bibnamefont {Ostrovskaya}},\ }\href
  {https://doi.org/10.1103/PhysRevLett.120.065301} {\bibfield  {journal}
  {\bibinfo  {journal} {Phys. Rev. Lett.}\ }\textbf {\bibinfo {volume} {120}},\
  \bibinfo {pages} {065301} (\bibinfo {year} {2018})}\BibitemShut {NoStop}%
\bibitem [{\citenamefont {Hutchison}\ \emph {et~al.}(2012)\citenamefont
  {Hutchison}, \citenamefont {Schwartz}, \citenamefont {Genet}, \citenamefont
  {Devaux},\ and\ \citenamefont {Ebbesen}}]{hutch2012}%
  \BibitemOpen
  \bibfield  {author} {\bibinfo {author} {\bibfnamefont {J.~A.}\ \bibnamefont
  {Hutchison}}, \bibinfo {author} {\bibfnamefont {T.}~\bibnamefont {Schwartz}},
  \bibinfo {author} {\bibfnamefont {C.}~\bibnamefont {Genet}}, \bibinfo
  {author} {\bibfnamefont {E.}~\bibnamefont {Devaux}},\ and\ \bibinfo {author}
  {\bibfnamefont {T.~W.}\ \bibnamefont {Ebbesen}},\ }\href
  {https://doi.org/10.1002/anie.201107033} {\bibfield  {journal} {\bibinfo
  {journal} {Angew. Chem.}\ }\textbf {\bibinfo {volume} {51}},\ \bibinfo
  {pages} {1592} (\bibinfo {year} {2012})}\BibitemShut {NoStop}%
\bibitem [{\citenamefont {Herrera}\ and\ \citenamefont
  {Spano}(2016)}]{herrera2016}%
  \BibitemOpen
  \bibfield  {author} {\bibinfo {author} {\bibfnamefont {F.}~\bibnamefont
  {Herrera}}\ and\ \bibinfo {author} {\bibfnamefont {F.~C.}\ \bibnamefont
  {Spano}},\ }\href {https://doi.org/10.1103/PhysRevLett.116.238301} {\bibfield
   {journal} {\bibinfo  {journal} {Phys. Rev. Lett.}\ }\textbf {\bibinfo
  {volume} {116}},\ \bibinfo {pages} {238301} (\bibinfo {year}
  {2016})}\BibitemShut {NoStop}%
\bibitem [{\citenamefont {Flick}\ \emph {et~al.}(2017)\citenamefont {Flick},
  \citenamefont {Ruggenthaler}, \citenamefont {Appel},\ and\ \citenamefont
  {Rubio}}]{Flick2017}%
  \BibitemOpen
  \bibfield  {author} {\bibinfo {author} {\bibfnamefont {J.}~\bibnamefont
  {Flick}}, \bibinfo {author} {\bibfnamefont {M.}~\bibnamefont {Ruggenthaler}},
  \bibinfo {author} {\bibfnamefont {H.}~\bibnamefont {Appel}},\ and\ \bibinfo
  {author} {\bibfnamefont {A.}~\bibnamefont {Rubio}},\ }\href
  {http://www.pnas.org/content/early/2017/03/07/1615509114} {\bibfield
  {journal} {\bibinfo  {journal} {Proceedings of the National Academy of
  Sciences}\ }\textbf {\bibinfo {volume} {114}},\ \bibinfo {pages} {3026}
  (\bibinfo {year} {2017})}\BibitemShut {NoStop}%
\bibitem [{\citenamefont {Ribeiro}\ \emph {et~al.}(2018)\citenamefont
  {Ribeiro}, \citenamefont {Martínez-Martínez}, \citenamefont {Du},
  \citenamefont {Campos-Gonzalez-Angulo},\ and\ \citenamefont
  {Yuen-Zhou}}]{Ribeiro2018}%
  \BibitemOpen
  \bibfield  {author} {\bibinfo {author} {\bibfnamefont {R.~F.}\ \bibnamefont
  {Ribeiro}}, \bibinfo {author} {\bibfnamefont {L.~A.}\ \bibnamefont
  {Martínez-Martínez}}, \bibinfo {author} {\bibfnamefont {M.}~\bibnamefont
  {Du}}, \bibinfo {author} {\bibfnamefont {J.}~\bibnamefont
  {Campos-Gonzalez-Angulo}},\ and\ \bibinfo {author} {\bibfnamefont
  {J.}~\bibnamefont {Yuen-Zhou}},\ }\href {https://doi.org/10.1039/C8SC01043A}
  {\bibfield  {journal} {\bibinfo  {journal} {Chem. Sci.}\ }\textbf {\bibinfo
  {volume} {9}},\ \bibinfo {pages} {6325} (\bibinfo {year} {2018})}\BibitemShut
  {NoStop}%
\bibitem [{\citenamefont {Thomas}\ \emph
  {et~al.}(2019{\natexlab{b}})\citenamefont {Thomas}, \citenamefont
  {Lethuillier-Karl}, \citenamefont {Nagarajan}, \citenamefont {Vergauwe},
  \citenamefont {George}, \citenamefont {Chervy}, \citenamefont {Shalabney},
  \citenamefont {Devaux}, \citenamefont {Genet}, \citenamefont {Moran},\ and\
  \citenamefont {Ebbesen}}]{Thomas2019}%
  \BibitemOpen
  \bibfield  {author} {\bibinfo {author} {\bibfnamefont {A.}~\bibnamefont
  {Thomas}}, \bibinfo {author} {\bibfnamefont {L.}~\bibnamefont
  {Lethuillier-Karl}}, \bibinfo {author} {\bibfnamefont {K.}~\bibnamefont
  {Nagarajan}}, \bibinfo {author} {\bibfnamefont {R.~M.~A.}\ \bibnamefont
  {Vergauwe}}, \bibinfo {author} {\bibfnamefont {J.}~\bibnamefont {George}},
  \bibinfo {author} {\bibfnamefont {T.}~\bibnamefont {Chervy}}, \bibinfo
  {author} {\bibfnamefont {A.}~\bibnamefont {Shalabney}}, \bibinfo {author}
  {\bibfnamefont {E.}~\bibnamefont {Devaux}}, \bibinfo {author} {\bibfnamefont
  {C.}~\bibnamefont {Genet}}, \bibinfo {author} {\bibfnamefont
  {J.}~\bibnamefont {Moran}},\ and\ \bibinfo {author} {\bibfnamefont {T.~W.}\
  \bibnamefont {Ebbesen}},\ }\href {https://doi.org/10.1126/science.aau7742}
  {\bibfield  {journal} {\bibinfo  {journal} {Science}\ }\textbf {\bibinfo
  {volume} {363}},\ \bibinfo {pages} {615} (\bibinfo {year}
  {2019}{\natexlab{b}})}\BibitemShut {NoStop}%
\bibitem [{\citenamefont {K\'{e}na-Cohen}\ and\ \citenamefont
  {Yuen-Zhou}(2019)}]{cohen2019}%
  \BibitemOpen
  \bibfield  {author} {\bibinfo {author} {\bibfnamefont {S.}~\bibnamefont
  {K\'{e}na-Cohen}}\ and\ \bibinfo {author} {\bibfnamefont {J.}~\bibnamefont
  {Yuen-Zhou}},\ }\href {https://doi.org/10.1021/acscentsci.9b00219} {\bibfield
   {journal} {\bibinfo  {journal} {ACS Cent. Sci.}\ }\textbf {\bibinfo {volume}
  {5}},\ \bibinfo {pages} {386} (\bibinfo {year} {2019})}\BibitemShut {NoStop}%
\end{thebibliography}%

\appendix

\begin{widetext}

\section{Validity of adiabatic approximation}
\label{sec:validity_class_elim}

In this appendix, we discuss in more detail the conditions required for the validity of the adiabatic elimination of $B$, using the derivation of the model in the classical limit (Sec.~\ref{sec:class_elim}). We derive the general condition, under which the steady-state solution for the $B$ ensemble [Eq.~\eqref{st_st_sol}] can be inserted in Eqs.~(\ref{Eq1_sys}) and (\ref{Eq2_sys}) to obtain the set of equations in~(\ref{sys_formal}). 

We first write Eq.~(\ref{sys}) as $\partial_{t} \vec{\beta} = {\bf B} \vec{\beta}(t) + \vec{S} (t)$ with ${\bf B} = - \mi {\bf M}$ and $\vec{S} (t) = -\mi \left[ \vec{G} \alpha(t) + \vec{V} \beta_A (t) \right]$. The formal solution of this differential equation reads $\vec{\beta} (t) = e^{{\bf B} t} \vec{\beta} (0) + \int_{0}^{t} {\rm d} \tau' e^{{\bf B} \tau'} \vec{S} (t -\tau')$. For simplicity, we assume the limit of large $t$ and set the upper limit of integration to infinity
\begin{align}
\vec{\beta} (t) &= \int_{0}^{\infty} {\rm d} \tau' e^{{\bf B} \tau'} \vec{S} (t -\tau').
\label{pro_act_tra}
\end{align}
The first term $e^{{\bf B} t} \vec{\beta} (0)$ vanishes when assuming the initial condition $\vec{\beta}(0)=0$ similarly as in Sec.~\ref{sec:full_elim}, or alternatively in the limit of large $t \gg |\Re \{ \lambda_j \}|^{-1}$ (with $\Re \{ \lambda_j \} < 0$), which leads to the damping of the initial condition. Here, $\lambda_j$ denote the eigenvalues of the matrix ${\bf B}$ [as in Sec.~\ref{sec:pert_2ord}]. These eigenvalues determine the fast dynamics of the interacting $B$ ensemble. The first term $e^{{\bf B} t} \vec{\beta} (0)$ becomes negligible when one averages over a time interval $\Delta t \gg |\lambda_j|^{-1}$, where $\Delta t$ is a coarse-grained time scale assumed to be large compared to the typical time scale associated to the dynamics of the $B$ ensemble, but small compared to the time scale of the effective dynamics of the subsystem $\mathcal{S}$. In other words, this requires a time-scale separation between the $B$ ensemble and the subsystem $\mathcal{S}$.   

Using the definition ${\bf B}= \sum_{j}\lambda_{j} \vec{x}_{j}\vec{x}^{\rm T}_{j}$ from Sec.~\ref{sec:pert_2ord}, Eq.~(\ref{pro_act_tra}) provides $\vec{\beta} (t) = \sum_{j} \vec{x}_{j} \int_{0}^{\infty} {\rm d} \tau' e^{\lambda_{j} \tau'} \vec{x}^{\rm T}_{j} \vec{S} (t -\tau')$. After integration by parts, the previous equation can be written as $\vec{\beta} (t) = \vec{\beta}_{\rm ad} (t) + \vec{\beta}_{\rm ret} (t)$ with
\begin{align}
\vec{\beta}_{\rm ad} (t) 
&= - \sum_{j} \frac{\vec{x}_{j}}{\lambda_{j}} \vec{x}^{\rm T}_{j}\vec{S} (t) \label{eq:sol_ad} 
= \mi \sum_{j} \frac{\vec{x}_{j}}{\lambda_{j}} \left[ \vec{x}^{\rm T}_{j}\vec{G} \alpha(t) + \vec{x}^{\rm T}_{j}\vec{V} \beta_A (t) \right], 
\\
\vec{\beta}_{\rm ret} (t) &= - \sum_j \frac{\vec{x}_{j}}{\lambda_{j}} \int_{0}^{\infty} {\rm d} \tau' e^{\lambda_j \tau'} \partial_{\tau'} \vec{x}^{\rm T}_{j} \vec{S} (t -\tau').& 
\label{eq:sol_ret}
\end{align}

Using the relation ${\bf B}^{-1} = \mi {\bf M}^{-1}= \sum_{j} \vec{x}_{j}\vec{x}^{\rm T}_{j} / \lambda_{j}$, one can write the {\it adiabatic solution} $\vec{\beta}_{\rm ad}(t)$ in Eq.~\eqref{eq:sol_ad} in the form of Eq.~(\ref{st_st_sol}), namely $\vec{\beta}_{\rm ad}(t) = - {\bf M}^{-1} \vec{G} \alpha(t)- {\bf M}^{-1} \vec{V}\beta_A(t)$. 
Furthermore, since $\alpha (t)$ and $\beta_{A} (t)$ are respectively of order $\sqrt{\bar{n}}$ (with $\bar{n}$ the mean photon number) and unity, the condition $|\vec{\beta} (t)|^2 \ll 1$ of low population for the spins $B$ requires $| \vec{x}_j^{\rm T} \vec{V} | \ll |\lambda_j|$ and $| \vec{x}_j^{\rm T} \vec{G} | \sqrt{\bar{n}} \ll |\lambda_j|$. In the limit of small $\bar{n}$ (vacuum-Rabi couplings), the latter requirement reads $| \vec{x}_j^{\rm T} \vec{G} | \ll |\lambda_j|$. The previous conditions are consistent with assuming a sufficiently weak coupling between the $B$ ensemble and the subsystem $\mathcal{S}$, which justifies a perturbative treatment of the interaction (as in Sec.~\ref{sec:eff_qme}).

The solution $\vec{\beta}_{\rm ret} (t)$ [Eq.~\eqref{eq:sol_ret}] is associated to retardation effects and was neglected in Eq.~\eqref{st_st_sol} provided $\vec{\beta}_{\rm ret} (t) \ll \vec{\beta}_{\rm ad} (t)$. In order to justify this approximation, we now estimate the contribution $\vec{\beta}_{\rm ret} (t)$ in a self-consistent manner, which allows to obtain the leading-order correction to the adiabatic solution. We start from Eq.~(\ref{eq:sol_ret}) with the definition of $\vec{S}(t)$, and calculate the time derivatives $\partial_{\tau'}\alpha (t-\tau')$ and $\partial_{\tau'}\beta_{A} (t-\tau')$ using Eqs.~(\ref{Eq1_sys}) and (\ref{Eq2_sys}). This leads to

\begin{align*}
\vec{\beta}_{\text{ret}} (t) = -\sum_j \frac{\vec{x}_{j}}{\lambda_j} \int_{0}^{\infty} {\rm d} \tau' e^{\lambda_j \tau'} \bigg\{& \vec{x}_{j}^{\rm T}\vec{G} \left[ \left( \Delta_c - \mi \kappa \right) \alpha (t-\tau') + g_A \beta_A (t-\tau') + \vec{G}^{\rm T} \vec{\beta} (t-\tau') \right] \\
&+ \vec{x}_{j}^{\rm T} \vec{V} \left[ -\mi \gamma_A \beta_A (t-\tau') + g_A \alpha(t-\tau') + \vec{V}^{\rm T} \vec{\beta}(t-\tau') \right]  \bigg\}.
\end{align*}
Now replacing $\vec{\beta}(t-\tau')$ by its adiabatic solution $\vec{\beta}(t-\tau') \approx \vec{\beta}_{\rm ad} (t-\tau')$, the term $\vec{\beta}_{\rm ret}(t)$ is estimated as
\begin{align*}
\vec{\beta}_{\text{ret}} (t) \approx -\sum_j \frac{\vec{x}_{j}}{\lambda_j} \int_{0}^{\infty} {\rm d} \tau' e^{\lambda_j \tau'} \bigg\{& \vec{x}_{j}^{\rm T}\vec{G} \Big[ \left( \Delta_c - \mi \kappa \right) \alpha (t) + g_A \beta_A (t) + \vec{G}^{\rm T} \Big(- {\bf M}^{-1} \vec{G} \alpha(t) - {\bf M}^{-1} \vec{V} \beta_A(t) \Big) \Big] \\ 
& +  \vec{x}_{j}^{\rm T} \vec{V} \Big[ -\mi \gamma_A \beta_A (t) + g_A \alpha (t) + \vec{V}^{\rm T} \Big( - {\bf M}^{-1} \vec{G} \alpha(t) - {\bf M}^{-1} \vec{V} \beta_A(t) \Big) \Big]  \bigg\}. 
\end{align*}
Here, we have also replaced $\alpha(t-\tau')$ by $\alpha(t)$ and $\beta_A(t-\tau')$ by $\beta_A(t)$ in the integrand similarly to the approximations $\rho_{gg}(t-\tau) \approx \rho_{gg}(t)$ and $\exp[(\mathcal{L}_B + \mathcal{L}_S) \tau] \approx \exp[\mathcal{L}_B \tau]$ of Sec.~\ref{sec:full_elim}, which hold when the time scales associated to the dynamics of $B$ and the subsystem $\mathcal{S}$ are well separated. Using the definitions of the effective parameters in Eq.~(\ref{eq:meq_parameters}), the retarded solution reads
\begin{align}
\vec{\beta}_{\text{ret}} (t) \approx \sum_j \frac{\vec{x}_{j}}{\lambda^{2}_{j}} \bigg\{  \vec{x}_{j}^{\rm T} \vec{G} \left[ \left(\Delta^{\text{eff}}_c - \mi \kappa^{\text{eff}}\right) \alpha (t) + \left(g^{\text{eff}}_A - \mi \mu \right)\beta_A (t)\right] +\vec{x}_{j}^{\rm T} \vec{V} \left[ \left(\Delta^{\text{eff}}_A -\mi \gamma^{\text{eff}}_A \right) \beta_A (t) + \left(g^{\text{eff}}_A - \mi \mu \right) \alpha (t)  \right]  \bigg\}.
\label{ret_eh_sol}
\end{align}

By comparing Eqs.~(\ref{eq:sol_ad}) and (\ref{ret_eh_sol}), we find that $\vert \Delta^{\text{eff}}_c - \mi \kappa^{\text{eff}} \vert \ll \vert \lambda_{j} \vert$, $\vert \Delta^{\text{eff}}_A -\mi \gamma^{\text{eff}}_A \vert \ll \vert \lambda_{j} \vert$ and $\sqrt{\bar{n}} \vert g^{\text{eff}}_A - \mi \mu \vert \ll \vert \lambda_{j} \vert$ are necessary conditions to neglect retardation effects ($\vec{\beta}_{\text{ret}} \ll \vec{\beta}_{\text{ad}}$). These conditions correspond to a separation of the different time scales, namely that the dynamics of the subsystem $\mathcal{S}$ is slow compared to that of $B$, consistently with the substitutions $\alpha(t-\tau') \approx \alpha(t)$ and $\beta_A(t-\tau') \approx \beta_A(t)$. 

In conclusion, the global condition for adiabatic elimination of $B$ is that $\lambda_{j}$ are the largest parameters of the problem, in agreement with the arguments of Sec.~\ref{sec:eff_qme} used in the derivation of the effective master equation. Note that when $B$ is reduced to a single spin, this condition becomes $\vert \Delta_{B} -\mi \gamma_{B} \vert \gg \left\{ |\Omega_{AB} - \mi \gamma_{AB}|, |g_B| \sqrt{\bar{n}} , |\Delta_c^{\rm eff} - \mi \kappa^{\rm eff}|, |\Delta_A^{\rm eff} - \mi \gamma_A^{\rm eff}| , \sqrt{\bar{n}} |g^{\text{eff}}_A - \mi \mu| \right\}$.

\end{widetext}

\end{document}